\documentclass{caosp308}

\usepackage{graphicx}

\usepackage{natbib}
\usepackage{longtable}
\articleNo{}
\pubyear{}
\volume{}
\volnumber{}
\firstpage{}
\received{September 26, 2021}
\accepted{October 22, 2021}

\def\BibTeX{{\rm B\kern-.05em{\sc i\kern-.025em b}\kern-.08em
             T\kern-.1667em\lower.7ex\hbox{E}\kern-.125emX}}

\begin{document}

%
\htitle{Si abundances of upper main-sequence stars}
\hauthor{Y. Takeda}

\title{Photospheric silicon abundances of \\
upper main-sequence stars derived from \\
Si~{\sc ii} 6347/6371 doublet lines}


%
%
\author{
        Y. Takeda\orcid{0000-0002-7363-0447}
       }

%
\institute{
          11-2 Enomachi, Naka-ku, Hiroshima-shi, 730-0851, Japan \\
          \email{ytakeda@js2.so-net.ne.jp}
          }

\date{}

\maketitle

\begin{abstract}
Silicon abundances were determined by applying the spectrum-fitting technic 
to the Si~{\sc ii} doublet lines at 6347 and 6371~\AA\ for a sample of 120 
main-sequence stars in the $T_{\rm eff}$ range of $\sim$~7000--14000~K 
(comprising not only normal stars but also non-magnetic chemically peculiar 
stars) with an aim of investigating their behaviors (e.g., correlation with
stellar parameters and abundances of other elements such as Fe or C) and 
the background physical mechanisms involved therein, where attention was 
paid to taking into account of the non-LTE effect and to assigning a reasonable 
value of microturbulence. The following trends were revealed from the analysis:
(i) The resulting [Si/H] values, mostly ranging from $\sim -0.5$ 
to $\sim +0.3$, show a positive correlation with [Fe/H].
(ii) A kind of anti-correlation exists between Si and C as seen from
the tendency of [C/Si] steeply decreasing with [Si/H].
(iii) Si abundances do not show any clear dependence upon $T_{\rm eff}$ or 
$v_{\rm e}\sin i$, while Am and HgMn stars appear to show comparatively higher 
[Si/H] than normal stars.
Although it is not straightforward to explain these observational facts, 
different physical processes (gas--dust separation and atomic diffusion) 
are likely to be intricately involved in producing these characteristic 
behaviors of Si composition in the surface of late A through late B dwarfs.
\vspace{0.1cm}
\end{abstract}

\section{Introduction}

It is known that a significant fraction of late A through late B-type 
main sequence stars show anomalous spectra indicative of surface
abundance anomalies. Those chemically peculiar (CP) stars are
divided into several groups according to their features
as summarized in the review paper by Preston (1974).
So far, the abundance characteristics of many elements in CP stars 
have been investigated in comparison with normal stars to discuss 
the origin of their anomalies. 

However, regarding silicon, an important abundant element used as 
the fiducial reference in geo- or cosmo-chemistry, its abundance 
behavior in upper main-sequence stars is not yet well understood.
While conspicuous overabundance of Si is known to be observed in 
magnetic CP stars (CP2; the group 2 in CP stars classified by
Preston 1974), how it behaves in non-magnetic CP stars (CP1 ---
Am stars; CP3 --- HgMn stars) or in normal stars is not clear.
As a matter of fact, we still do not know whether any Si anomaly
ever exists in these stars. According to the atomic diffusion 
theory (see, e.g., Michaud et al. 2015),
which is considered to be a promising mechanism to explain the 
origin of abundance characteristics in CP stars, Si is expected to
be somewhat underabundant (Richer et al. 2000; Talon et al. 2006).
Meanwhile, such a trend is not necessarily seen in spectroscopically 
determined Si abundances of A- and late B-type dwarfs, which are rather 
diversified around the normal (solar) abundance (somewhat overabundant 
or underabundant depending on cases; e.g., Niemczura et al. 2015; 
Ghazaryan, Alecian 2016; Mashonkina et al. 2020b; Saffe et al. 2021), 
though Si abundance determination significantly depends upon which 
line is to be used.

Another noteworthy aspect characterizing the importance of Si 
abundance is that this element is a typical refractory species (being 
easily fractionated into dust) in contrast to the volatile elements
such as C, N, and O. Interestingly, Holweger and St\"{u}renburg (1993)
reported that  even normal early A-type stars (like $\lambda$ Boo-type stars) 
show anti-correlation between the abundances of Si and C ([C/Si] systematically 
decreases with [Si/H]), which means that some kind of gas--dust separation 
process (its degree being different from star to star) would have operated 
in the star formation phase and altered the primordial composition of gas. 
Is such an effect observed also stars of other types (i.e., late A through 
late B stars including CP1 and CP3 stars)? This is an interesting problem 
to be clarified.

Conveniently, Takeda et al. (2018; hereinafter referred to as T18) recently
published the C, N, and O abundances for a large sample of 100 main-sequence
stars (comprising normal as well as non-magnetic CP stars) covering 
7000~$\la T_{\rm eff} \la$~11000~K. It would be worthwhile, therefore,
to  determine the Si abundances for these stars. 
This would enable to clarify the behaviors of both [Si/H] and [C/Si], 
by which the nature of abundance peculiarity of Si (if any exists) 
in late A through late B-type stars and the involved physical process 
may be investigated. This is the aim of the present study. 

In the past Si abundance determinations in upper main-sequence stars so far,
it appears that neutral Si~{\sc i} lines were mainly used in late--mid A-type 
stars (including classical Am stars) while once-ionized Si~{\sc ii} lines were 
primarily employed in early A and late B stars (because Si~{\sc i} lines 
quickly fade out with an increase in $T_{\rm eff}$). 
Since the mixed use of lines of different ionization stages is not advantageous
because of inevitable line-by-line abundance discrepancies (see, e.g., 
Mashonkina 2020a), we decided to invoke in this study only the Si~{\sc ii} 
doublet lines at 6347 and 6371~\AA, which are of high quality (i.e., almost 
free from blending) and sufficiently strong over the whole relevant 
$T_{\rm eff}$ range. Nevertheless, some disadvantages are involved in using
these strong Si~{\sc ii} lines; that is, the resulting Si abundances suffer 
an appreciable non-LTE affect and are sensitive to the microturbulence 
parameter. Accordingly, special attention had to be paid to these two points.

\section{Observational data}

Regarding the program stars in this study, all the 101 targets (including
the reference star Procyon) in T18 (cf. Section~2 therein) were 
adopted without change. In addition, in order to back up the range of 
11000~$\la T_{\rm eff} \la$~14000~K (which was not covered in T18), 
19 late B-type stars (among which $\sim$~40\% are CP3 stars) were newly included.
As such, our targets are 120 late B-type through early F-type stars on or near 
to the main sequence (luminosity classes of III--V) which have slow to 
moderately-high rotational velocities
(0~km~s$^{-1}$~$\la v_{\rm e}\sin i \la 100$~km~s$^{-1}$). 
Among these, about $\sim 1/3$ are non-magnetic CP stars: 25 Am stars, 
13 HgMn (or Mn) stars, and 2 $\lambda$~Boo stars. Besides, 
our sample includes 16 Hyades A-type stars.
The list of these 120 stars is given in Table~1, while the data source 
and the basic information of the observational materials are summarized 
in Table~2.

\begin{figure}
\centerline{\includegraphics[width=0.5\textwidth,clip=]{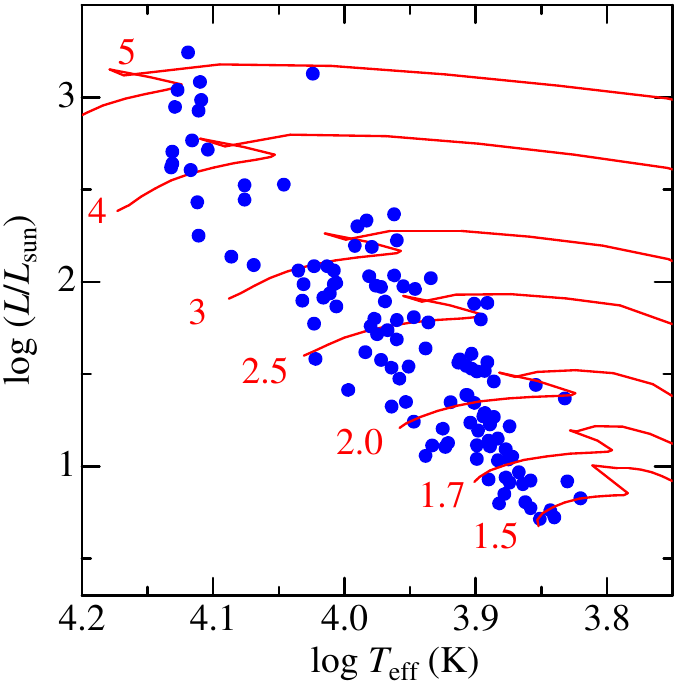}}
\caption{
The 120 program stars are plotted on the theoretical HR diagram ($\log (L/L_{\odot})$ 
vs. $\log T_{\rm eff}$), where $T_{\rm eff}$ was derived from colors (cf. Section~3) 
and $L$ was evaluated from visual magnitude (corrected for interstellar extinction 
by following Arenou et al. 1992), Hipparcos parallax (van Leeuwen 2007), and 
bolometric correction (Flower 1996). Theoretical solar-metallicity tracks for 
7 different masses (1.5, 1.7, 2, 2.5, 3, 4, and 5~$M_{\odot}$), which were computed 
by Lejeune and Schaerer (2001), are also depicted by solid lines for comparison.
}
\label{fig1}
\end{figure}

\setcounter{table}{0}
\begin{table}
\caption{Program stars and the results of the analysis.}
\tiny
\begin{center}
\begin{tabular}{cccccccccccccl}\hline
\hline
HD\# &  Name & Sp.Type & $T_{\rm eff}$ & $\log g$ & [Fe/H] & $v_{\rm e}\sin i$ & 
[C/H] & $\xi_{\rm std}$ & [Si/H]$_{\rm std}$ & $\xi_{\rm refit}$ & [Si/H]$_{\rm refit}$ & Group & Remark \\
(1) & (2) & (3) & (4) & (5) & (6) & (7) & (8) & (9) & (10) & (11) & (12) & (13) & (14) \\
\hline
002628 & 28 And              & A7III      &  7143 &  3.48 & $-$0.27 &  19 & $-$0.32 &  3.35 & $-$0.27 &  4.27 & $-$0.29 & D &            \\
005448 & $\mu$ And           & A5V        &  8147 &  3.82 & $-$0.14 &  72 & $-$0.32 &  3.98 & $-$0.31 &  1.31 & +0.22 & B &            \\
011529 & $\omega$ Cas        & B8III      & 12858 &  3.43 & $-$0.15 &  30 & $\cdots$ &  1.00 & $-$0.02 & $\cdots$ & $\cdots$ & F & SBo        \\
011636 & $\beta$ Ari         & A5V...     &  8294 &  4.12 & +0.15 &  73 & $-$0.49 &  3.93 & $-$0.09 &  4.01 & $-$0.02 & B & SBo        \\
012111 & 48 Cas              & A3IV       &  7910 &  4.08 & $-$0.23 &  76 & $-$0.46 &  3.99 & $-$0.27 &  1.26 & +0.25 & B & SBo        \\
012216 & 50 Cas              & A2V        &  9553 &  3.90 & +0.15 &  88 & $-$0.20 &  2.58 & $-$0.29 &  2.27 & $-$0.17 & B & SB2        \\
013161 & $\beta$ Tri         & A5III      &  7957 &  3.68 & $-$0.32 &  65 & $-$0.42 &  4.00 & $-$0.48 & $\cdots$ & $\cdots$ & B & SB2o       \\
014252 & 10 Tri              & A2V        &  9023 &  3.64 & $-$0.04 &  23 & $-$0.53 &  3.27 & $-$0.32 &  2.19 & $-$0.08 & D & V          \\
016350 &                     & B9.5V      &  9824 &  3.72 & $-$0.03 &  21 & $-$0.63 &  2.22 & $-$0.25 & $\cdots$ & $\cdots$ & D &            \\
017081 & $\pi$ Cet           & B7IV       & 13063 &  3.72 & $-$0.10 &  20 & $\cdots$ &  1.00 & +0.05 &  2.98 & $-$0.18 & F & SB         \\
017093 & 38 Ari              & A7III-IV   &  7541 &  3.95 & $-$0.23 &  69 & $-$0.46 &  3.81 & $-$0.26 & $\cdots$ & $\cdots$ & B & V          \\
018454 & 4 Eri               & A5IV/V     &  7740 &  4.07 & +0.24 & 100 & $-$0.40 &  3.94 & +0.07 & $\cdots$ & $\cdots$ & B & V          \\
020149 &                     & A1Vs       &  9522 &  3.99 & $-$0.05 &  21 & $-$0.25 &  2.62 & $-$0.31 &  2.63 & $-$0.24 & D & SB?        \\
020320 & $\zeta$ Eri         & A5m        &  7505 &  3.91 & $-$0.12 &  67 & $-$0.72 &  3.78 & $-$0.30 &  2.36 & $-$0.03 & B & SBo, Am    \\
020346 &                     & A2IV       &  8824 &  3.56 & +0.07 &  21 & $-$0.47 &  3.50 & $-$0.28 &  2.98 & $-$0.13 & D & SB?        \\
023281 &                     & A5m        &  7761 &  4.19 & +0.05 &  76 & $-$0.36 &  3.95 & $-$0.16 &  3.14 & +0.02 & B & Am    \\
023408 & 20 Tau              & B8III      & 12917 &  3.36 & +0.16 &  30 & $\cdots$ &  1.00 & $-$0.74 &  1.39 & $-$0.71 & F & SB         \\
023878 & $\tau^{7}$ Eri      & A1V        &  8674 &  3.80 & +0.18 &  25 & $-$0.63 &  3.65 & $-$0.17 &  3.83 & $-$0.12 & D & V?         \\
024740 & 32 Tau              & F2IV       &  6768 &  3.77 & $-$0.11 &  21 & $-$0.07 &  2.71 & +0.00 &  4.58 & $-$0.12 & D & V          \\
025490 & $\nu$ Tau           & A1V        &  9077 &  3.93 & $-$0.05 &  82 & $-$0.37 &  3.20 & $-$0.41 &  3.19 & $-$0.33 & B &            \\
026322 & 44 Tau              & F2IV-V     &  6795 &  3.46 & $-$0.15 &   6 & $-$0.14 &  2.76 & $-$0.20 &  2.68 & $-$0.12 & D &            \\
027045 & $\omega^{2}$ Tau    & A3m        &  7552 &  4.26 & +0.36 &  62 & $-$0.89 &  3.82 & +0.10 &  3.63 & +0.20 & B & SB, Am    \\
027628 & 60 Tau              & A3m        &  7218 &  4.05 & +0.10 &  30 & $-$1.08 &  3.45 & +0.15 &  4.26 & +0.12 & B & SB1o, Am, H \\
027749 & 63 Tau              & A1m        &  7448 &  4.21 & +0.41 &  13 & $-$1.30 &  3.73 & +0.09 &  4.05 & +0.12 & B & SB1o, Am, H \\
027819 & $\delta^{2}$ Tau    & A7V        &  8047 &  3.95 & $-$0.05 &  45 & $-$0.15 &  4.00 & $-$0.17 &  4.33 & $-$0.14 & B & SB, H \\
027934 & $\kappa^{1}$ Tau    & A7IV-V     &  8159 &  3.84 & +0.02 &  83 & $-$0.17 &  3.98 & $-$0.21 &  3.70 & $-$0.10 & B & SB?, H \\
027962 & $\delta^{3}$ Tau    & A2IV       &  8923 &  3.94 & +0.25 &  11 & $-$0.49 &  3.39 & $-$0.08 &  3.08 & +0.03 & A & SB, H \\
028226 &                     & Am         &  7361 &  4.01 & +0.31 &  81 & $-$0.37 &  3.63 & +0.16 & $\cdots$ & $\cdots$ & B & SB2, Am, H \\
028319 & $\theta^{2}$ Tau    & A7III      &  7789 &  3.68 & $-$0.13 &  68 & $-$0.26 &  3.96 & $-$0.22 &  3.28 & $-$0.05 & B & SB1o, H \\
028355 & 79 Tau              & A7V        &  7809 &  3.98 & +0.19 &  87 & $-$0.39 &  3.97 & $-$0.06 &  2.37 & +0.27 & B & V?, H \\
028546 & 81 Tau              & Am         &  7640 &  4.17 & +0.23 &  28 & $-$0.44 &  3.88 & $-$0.05 &  4.18 & $-$0.01 & B & V?, Am, H \\
029388 & 90 Tau              & A6V        &  8194 &  3.88 & $-$0.01 &  82 & $-$0.18 &  3.97 & $-$0.13 &  3.28 & +0.05 & B & SB1, H \\
029479 & $\sigma^{1}$ Tau    & A4m        &  8406 &  4.14 & +0.35 &  56 & $-$0.52 &  3.87 & +0.06 &  4.06 & +0.10 & B & SBo, Am, H \\
029499 &                     & A5m        &  7638 &  4.08 & +0.29 &  61 & $-$0.44 &  3.88 & +0.19 &  3.98 & +0.25 & B & V, Am, H \\
030121 & 4 Cam               & A3m        &  7700 &  3.98 & +0.27 &  57 & $\cdots$ &  3.92 & +0.03 &  3.34 & +0.19 & B & Am    \\
030210 &                     & Am...      &  7927 &  3.94 & +0.40 &  56 & $-$1.60 &  4.00 & +0.19 &  4.46 & +0.20 & B & SB1?, Am, H \\
032115 &                     & A8IV       &  7207 &  4.13 & $-$0.06 &  10 & $-$0.11 &  3.44 & $-$0.10 &  3.90 & $-$0.06 & D & V          \\
032537 & 9 Aur               & F0V        &  6970 &  4.07 & $-$0.18 &  20 & $-$0.26 &  3.07 & $-$0.19 &  4.64 & $-$0.23 & D & SBo        \\
033204 &                     & A5m        &  7530 &  4.06 & +0.18 &  34 & $-$0.85 &  3.80 & +0.01 &  3.12 & +0.18 & B & Am, H \\
033254 & 16 Ori              & A2m        &  7747 &  4.14 & +0.28 &  13 & $-$1.33 &  3.94 & $-$0.02 &  3.90 & +0.06 & B & SBo, Am, H \\
033641 & $\mu$ Aur           & A4m        &  7961 &  4.21 & +0.18 &  79 & $-$0.49 &  4.00 & $-$0.07 &  0.82 & +0.57 & B & V, Am    \\
038899 & 134 Tau             & B9IV       & 10774 &  4.02 & +0.00 &  26 & $\cdots$ &  1.16 & $-$0.05 &  1.69 & $-$0.05 & F & V          \\
039945 &                     & A5V        &  7827 &  3.36 & $-$0.33 &  22 & $-$0.34 &  3.97 & $-$0.52 &  2.82 & $-$0.31 & D &            \\
040626 &                     & B9.5IV     & 10263 &  4.00 & +0.20 &  14 & $-$0.26 &  1.69 & $-$0.08 &  1.12 & +0.08 & D &            \\
040932 & $\mu$ Ori           & Am...      &  8005 &  3.93 & $-$0.12 &  11 & $-$0.64 &  4.00 & $-$0.35 &  3.45 & $-$0.22 & B & SB1o, Am, H \\
042035 &                     & B9V        & 10575 &  3.82 & $-$0.16 &   2 & $-$0.68 &  1.35 & $-$0.56 &  1.09 & $-$0.47 & D & V          \\
043378 & 2 Lyn               & A2Vs       &  9210 &  4.09 & $-$0.15 &  46 & $-$0.27 &  3.03 & $-$0.32 &  2.99 & $-$0.25 & B & V?         \\
045394 & 16 Gem              & A2Vs       &  8630 &  3.42 & $-$0.40 &  23 & $-$0.62 &  3.69 & $-$0.51 &  3.16 & $-$0.38 & D &            \\
047105 & $\gamma$ Gem        & A0IV       &  9115 &  3.49 & $-$0.03 &  11 & $-$0.27 &  3.16 & $-$0.39 &  2.58 & $-$0.24 & A & SB         \\
048915 & $\alpha$ CMa        & A0m...     &  9938 &  4.31 & +0.45 &  17 & $-$1.09 &  2.08 & +0.06 &  1.82 & +0.17 & A & SBo, Am    \\
053244 & $\gamma$ CMa        & B8II       & 13467 &  3.42 & +0.06 &  36 & $\cdots$ &  1.00 & $-$0.13 & $\cdots$ & $\cdots$ & F & V          \\
054834 &                     & A9V        &  7273 &  4.21 & +0.03 &  30 & $-$0.20 &  3.53 & $-$0.13 &  3.32 & $-$0.03 & D &            \\
058142 & 21 Lyn              & A1V        &  9384 &  3.74 & $-$0.05 &  19 & $-$0.45 &  2.81 & $-$0.30 &  1.36 & +0.01 & D & V          \\
060179 & $\alpha$ Gem        & A2Vm       &  9122 &  3.88 & $-$0.02 &  19 & $-$0.93 &  3.15 & $-$0.23 &  2.44 & $-$0.05 & B & SB1o, Am    \\
061421 & $\alpha$ CMi        & F5IV-V     &  6612 &  4.00 & +0.00 &   9 & +0.00 &  1.97 & +0.00 &  2.67 & +0.00 & E & SBo        \\
067959 &                     & A1V        &  9168 &  3.65 & +0.07 &  16 & $-$0.47 &  3.09 & $-$0.24 &  2.90 & $-$0.14 & D &            \\
072037 & 2 UMa               & A2m        &  7918 &  4.16 & +0.19 &  12 & $-$1.53 &  3.99 & $-$0.18 &  3.53 & $-$0.06 & B & Am    \\
072660 &                     & A1V        &  9635 &  3.97 & +0.37 &   5 & $-$0.77 &  2.47 & +0.02 &  2.31 & +0.11 & D &            \\
074198 & $\gamma$ Cnc        & A1IV       &  9381 &  4.11 & +0.25 &  85 & $-$0.30 &  2.81 & $-$0.09 &  2.45 & +0.04 & B & SB         \\
075469 &                     & A2Vs       &  9165 &  3.51 & $-$0.08 &  22 & $-$0.42 &  3.09 & $-$0.31 &  2.49 & $-$0.14 & D &            \\
076543 & o$^{1}$ Cnc         & A5III      &  8330 &  4.18 & +0.38 &  91 & $-$0.54 &  3.91 & +0.13 &  0.50 & +0.81 & B & SB         \\
077350 & $\nu$ Cnc           & A0III      & 10141 &  3.68 & +0.24 &  20 & $-$0.56 &  1.83 & $-$0.34 & $\cdots$ & $\cdots$ & C & SBo, Hg    \\
078316 & $\kappa$ Cnc        & B8IIIMNp   & 13513 &  3.85 & +0.33 &   8 & $\cdots$ &  1.00 & +0.02 & $\cdots$ & $\cdots$ & C & SB1o, Hg    \\
079158 & 36 Lyn              & B8IIIMNp   & 13535 &  3.72 & +1.05 &  46 & $\cdots$ &  1.00 & +0.32 &  3.25 & +0.03 & C & V, Hg    \\
079469 & $\theta$ Hya        & B9.5V      & 10510 &  4.20 & $-$0.02 &  82 & $-$0.30 &  1.42 & $-$0.49 &  1.19 & $-$0.40 & B & SB         \\
084107 & 15 Leo              & A2IV       &  8665 &  4.31 & +0.01 &  38 & $-$0.34 &  3.66 & $-$0.30 &  3.79 & $-$0.24 & B &            \\
089021 & $\lambda$ UMa       & A2IV       &  8861 &  3.61 & +0.08 &  52 & $-$0.57 &  3.46 & $-$0.16 &  3.18 & $-$0.05 & B & V          \\
089822 &                     & A0sp...    & 10307 &  3.89 & +0.47 &   5 & $-$0.45 &  1.64 & $-$0.07 &  1.49 & +0.02 & C & SB2o, Hg    \\
095382 & 59 Leo              & A5III      &  8017 &  3.95 & $-$0.09 &  68 & $-$0.29 &  4.00 & $-$0.17 &  3.71 & $-$0.06 & B &            \\
095418 & $\beta$ UMa         & A1V        &  9489 &  3.85 & +0.24 &  44 & $-$0.63 &  2.67 & $-$0.10 &  2.45 & +0.01 & B & SB         \\
095608 & 60 Leo              & A1m        &  8972 &  4.20 & +0.31 &  18 & $-$1.16 &  3.33 & $-$0.05 &  3.50 & +0.00 & B & Am    \\
098664 & $\sigma$ Leo        & B9.5Vs     & 10194 &  3.75 & $-$0.11 &  62 & $-$0.29 &  1.77 & $-$0.33 &  1.52 & $-$0.22 & C & SB         \\
106625 & $\gamma$ Crv        & B8III      & 11902 &  3.36 & $-$0.51 &  37 & $\cdots$ &  1.00 & $-$1.49 &  1.63 & $-$1.46 & C & SB, Hg    \\
116656 & $\zeta$ UMa         & A2V        &  9317 &  4.10 & +0.28 &  59 & $-$0.96 &  2.89 & $-$0.06 &  2.74 & +0.03 & B & SB2o       \\
129174 & $\pi^{1}$ Boo       & B9p MnHg   & 12929 &  4.02 & $-$0.06 &  16 & $\cdots$ &  1.00 & +0.05 &  0.94 & +0.13 & C & SB, Hg    \\
130557 &                     & B9Vsvar... & 10142 &  3.85 & +0.59 &  55 & $-$0.55 &  1.83 & $-$0.24 &  1.48 & $-$0.11 & C &            \\
130841 & $\alpha^{2}$ Lib    & A3IV       &  8079 &  3.96 & $-$0.24 &  58 & $-$1.60 &  3.99 & $-$0.63 & $\cdots$ & $\cdots$ & B & SB         \\
140436 & $\gamma$ CrB        & A1Vs       &  9274 &  3.89 & $-$0.27 &  68 & $-$1.34 &  2.95 & $-$0.70 &  1.61 & $-$0.49 & B &            \\
141795 & $\epsilon$ Ser      & A2m        &  8367 &  4.24 & +0.25 &  32 & $-$1.01 &  3.89 & $-$0.11 &  4.62 & $-$0.12 & B & V, Am    \\
143807 & $\iota$ CrB         & A0p...     & 10828 &  4.06 & +0.35 &   3 & $-$0.53 &  1.12 & +0.00 &  0.96 & +0.09 & D & SB, Hg    \\
144206 & $\upsilon$ Her      & B9III      & 11925 &  3.79 & +0.01 &  12 & $\cdots$ &  1.00 & $-$0.38 &  0.57 & $-$0.27 & C & Hg    \\
145389 & $\phi$ Her          & B9MNp...   & 11714 &  4.02 & +0.15 &  11 & $\cdots$ &  1.00 & $-$0.35 & $\cdots$ & $\cdots$ & C & SB1o, Hg    \\
149121 & 28 Her              & B9.5III    & 10748 &  3.89 & +0.24 &  10 & $-$0.73 &  1.19 & $-$0.40 &  0.97 & $-$0.31 & C & Hg    \\
150100 & 16 Dra              & B9.5Vn     & 10542 &  3.84 & $-$0.33 &  36 & +0.01 &  1.39 & $-$0.79 & $\cdots$ & $\cdots$ & C & V          \\
155763 & $\zeta$ Dra         & B6III      & 13397 &  4.24 & +0.07 &  41 & $\cdots$ &  1.00 & +0.09 &  0.88 & +0.18 & F & V          \\
158716 &                     & A1V        &  9214 &  4.30 & +0.28 &   4 & $-$0.57 &  3.03 & $-$0.01 &  2.79 & +0.09 & D &            \\
161701 &                     & B9V        & 12692 &  4.04 & +0.85 &  20 & $\cdots$ &  1.00 & $-$0.62 & $\cdots$ & $\cdots$ & C & SB2o, Hg    \\
172167 & $\alpha$ Lyr        & A0Vvar     &  9435 &  3.99 & $-$0.53 &  22 & $-$0.21 &  2.74 & $-$0.65 &  2.43 & $-$0.54 & A & V, LB    \\
173648 & $\zeta^{1}$ Lyr     & Am         &  8004 &  3.90 & +0.32 &  32 & $-$0.69 &  4.00 & +0.01 &  4.38 & +0.03 & B & SB1o, Am    \\
173880 & 111 Her             & A5III      &  8567 &  4.27 & +0.22 &  72 & $-$0.08 &  3.75 & $-$0.07 &  3.57 & +0.02 & B & SB?        \\
174567 &                     & A0Vs       &  9778 &  3.59 & +0.01 &  10 & $-$0.40 &  2.28 & $-$0.28 &  1.95 & $-$0.16 & D &            \\
176984 & 14 Aql              & A1V        &  9623 &  3.42 & +0.04 &  29 & $-$0.29 &  2.49 & $-$0.24 &  1.24 & +0.06 & D & V?         \\
179761 & 21 Aql              & B8II-III   & 12895 &  3.46 & $-$0.11 &  16 & $\cdots$ &  1.00 & $-$0.04 & $\cdots$ & $\cdots$ & F & V          \\
182564 & $\pi$ Dra           & A2IIIs     &  9125 &  3.80 & +0.39 &  27 & $-$0.35 &  3.14 & $-$0.03 &  3.59 & $-$0.03 & A &            \\
189849 & 15 Vul              & A4III      &  7870 &  3.62 & $-$0.08 &  11 & $-$0.17 &  3.99 & $-$0.36 &  4.09 & $-$0.30 & A & SBo        \\
\hline
\end{tabular}
\end{center}
\end{table}

\setcounter{table}{0}
\begin{table}
\caption{(Continued.)}
\tiny
\begin{center}
\begin{tabular}{cccccccccccccl}\hline
\hline
HD\# &  Name & Sp.Type & $T_{\rm eff}$ & $\log g$ & [Fe/H] & $v_{\rm e}\sin i$ & 
[C/H] & $\xi_{\rm std}$ & [Si/H]$_{\rm std}$ & $\xi_{\rm refit}$ & [Si/H]$_{\rm refit}$ & Group & Remark \\
(1) & (2) & (3) & (4) & (5) & (6) & (7) & (8) & (9) & (10) & (11) & (12) & (13) & (14) \\
\hline
190229 &                     & B9MNp...   & 13102 &  3.46 & +0.72 &  10 & $\cdots$ &  1.00 & $-$0.30 & $\cdots$ & $\cdots$ & C & SB1, Hg    \\
192640 & 29 Cyg              & A2V        &  8845 &  3.86 & $-$1.41 &  74 & +0.08 &  3.48 & $-$1.52 & $\cdots$ & $\cdots$ & B & V, LB    \\
193432 & $\nu$ Cap           & B9IV       & 10180 &  3.91 & +0.02 &  23 & $-$0.27 &  1.78 & $-$0.12 &  1.47 & +0.01 & D & V?         \\
193452 &                     & B9.5III/IV & 10543 &  4.15 & +0.39 &   3 & $-$0.86 &  1.39 & $-$0.09 &  1.24 & $-$0.01 & C & SB1o, Hg    \\
195725 & $\theta$ Cep        & A7III      &  7816 &  3.74 & +0.16 &  49 & $-$0.57 &  3.97 & $-$0.02 &  4.64 & $-$0.04 & B & SB2o       \\
196385 &                     & A9V        &  6919 &  4.23 & $-$0.21 &  15 & $-$0.17 &  2.98 & $-$0.14 &  4.13 & $-$0.15 & D &            \\
196426 &                     & B8IIIp     & 12899 &  3.89 & $-$0.10 &   6 & $\cdots$ &  1.00 & $-$0.01 & $\cdots$ & $\cdots$ & F &            \\
197392 &                     & B8II-III   & 13166 &  3.46 & $-$0.01 &  30 & $\cdots$ &  1.00 & $-$0.06 & $\cdots$ & $\cdots$ & C & SB         \\
198639 & 56 Cyg              & A4me...    &  7921 &  4.09 & +0.02 &  59 & $-$0.38 &  3.99 & $-$0.23 &  2.39 & +0.08 & B & V?, Am    \\
198667 & 5  Aqr              & B9III      & 11125 &  3.42 & +0.01 &  26 & $-$0.24 &  1.00 & $-$0.10 & $\cdots$ & $\cdots$ & C & V          \\
200499 & $\eta$ Cap          & A5V        &  8081 &  3.95 & $-$0.17 &  62 & $-$0.31 &  3.99 & $-$0.34 &  4.26 & $-$0.30 & B & V          \\
201433 &                     & B9V        & 12193 &  4.24 & +0.00 &  15 & $\cdots$ &  1.00 & +0.00 & $\cdots$ & $\cdots$ & C & SBo        \\
202671 & 30 Cap              & B5II/III   & 13566 &  3.36 & +0.45 &  25 & $\cdots$ &  1.00 & $-$0.22 & $\cdots$ & $\cdots$ & C & V?         \\
204188 &                     & A8m        &  7622 &  4.21 & +0.02 &  36 & $-$0.43 &  3.87 & $-$0.18 &  2.22 & +0.13 & B & SBo, Am    \\
207098 & $\delta$ Cap        & A5mF2 (IV) &  7312 &  4.06 & +0.21 &  81 & $-$2.00 &  3.57 & +0.02 & $\cdots$ & $\cdots$ & B & SBo, Am    \\
209625 & 32 Aqr              & A5m        &  7700 &  3.87 & +0.24 &   7 & $-$0.72 &  3.92 & +0.02 &  4.09 & +0.07 & D & SB1o, Am    \\
211236 &                     & A8/A9IV/V  &  7488 &  3.96 & $-$0.21 &  13 & $-$0.39 &  3.76 & $-$0.29 &  4.11 & $-$0.25 & D &            \\
212061 & $\gamma$ Aqr        & A0V        & 10384 &  3.95 & $-$0.08 &  54 & $-$0.49 &  1.55 & $-$0.42 &  2.48 & $-$0.46 & B & SB         \\
214994 & o Peg               & A1IV       &  9453 &  3.64 & +0.18 &   6 & $-$0.73 &  2.71 & $-$0.18 &  2.57 & $-$0.09 & A & V          \\
216627 & $\delta$ Aqr        & A3V        &  8587 &  3.59 & $-$0.25 &  79 & $-$0.45 &  3.73 & $-$0.63 &  3.87 & $-$0.57 & B & V          \\
218396 &                     & A5V        &  7091 &  4.06 & $-$0.59 &  41 & $-$0.11 &  3.27 & $-$0.49 &  3.54 & $-$0.44 & B &            \\
219485 &                     & A0V        &  9577 &  3.81 & $-$0.05 &  27 & $-$0.38 &  2.55 & $-$0.33 &  1.82 & $-$0.15 & D &            \\
222345 & $\omega^{1}$ Aqr    & A7IV       &  7487 &  3.88 & $-$0.07 &  86 & $-$0.32 &  3.76 & $-$0.18 &  3.11 & $-$0.02 & B & SB         \\
222603 & $\lambda$ Psc       & A7V        &  7757 &  3.99 & $-$0.17 &  56 & $-$0.27 &  3.95 & $-$0.25 &  2.58 & +0.02 & B & SB         \\
224995 & 31 Psc              & A6V        &  7779 &  3.64 & $-$0.13 &  99 & $-$0.23 &  3.96 & $-$0.26 & $\cdots$ & $\cdots$ & D & V          \\
\hline
\end{tabular}
\end{center}
(1) HD number. (2) Bayer/Flamsteed name. (3) Spectral type taken from Hipparcos catalogue (ESA 1997).
(4) Effective temperature (in K). (5) Logarithm of surface gravity ($\log g$ in dex,
where $g$ is in unit of cm~s$^{-2}$). (6) Fe abundance relative to Procyon.
(7) Projected rotational velocity (in km~s$^{-1}$). 
(8) Non-LTE carbon abundance relative to Procyon determined by Takeda et al. (2018).
(9) Standard microturbulent velocity (in km~s$^{-1}$) derived by Equation~(1). 
(10) Non-LTE silicon abundance relative to Procyon corresponding to $\xi_{\rm std}$.
(11) Directly determined microturbulence (in km~s$^{-1}$) as a result of spectrum refitting.
(12) Non-LTE silicon abundance relative to Procyon corresponding to $\xi_{\rm refit}$. 
(13) Group of the data source (cf. Table~2).
(14) Specific remark [spectroscopic binary (SB, ``o'' denotes the case where orbital elements 
are available) or radial velocity variable (V), chemical peculiarity type (Am or HgMn or 
$\lambda$~Boo), membership of Hyades cluster (H)]. The assigned CP classes were determined 
by consulting the spectral classifications in three sources: Hipparcos catalogue (ESA 1997), 
Bright Star Catalogue (Hoffleit, Jaschek 1991), and SIMBAD.  
\end{table}

\setcounter{table}{1}
\begin{table}
\scriptsize
\caption{Basic information of the observational data.}
\begin{center}
\begin{tabular}{ccccccc}\hline\hline
Group & $^{\#}$Instr. & Obs. Time & Resolution & Number & Star Type & Reference \\
\hline
$^{\dagger}$A & HIDES & 2008 Oct & 100000 & 7 & A type & Takeda et al. (2012) \\
B & BOES & 2008 Jan/Sep, 2009 Jan & 45000 & 56 & A type & Takeda et al. (2008, 2009) \\
C & HIDES & 2012 May  & 70000 & 8 & late B type & Takeda et al. (2014) \\
D & HIDES & 2017 Aug/Nov & 100000 & 29 & A type & Takeda et al. (2018)\\
E & HIDES & 2001 Feb & 70000 & 1 & Procyon & Takeda et al. (2005a) \\
F & HIDES & 2006 Oct & 70000 & 19 & late B type & Takeda et al. (2010) \\
\hline
\end{tabular} 
\end{center}
$^{\dagger}$Only for HD~172167 (Vega), Takeda et al.'s (2007) OAO/HIDES spectrum of high-S/N 
($\sim 2000$) and high-resolution ($\sim 100000$) observed in 2006 May was adopted.\\
$^{\#}$HIDES and BOES denote ``HIgh Dispersion Echelle Spectrograph'' at Okayama Astrophysical 
Observatory and ``Bohyunsan Observatory Echelle Spectrograph'' at Bohyunsan Optical Astronomy 
Observatory, respectively.
\end{table}

\section{Stellar parameters}

As in T18, the effective temperature ($T_{\rm eff}$) and the 
surface gravity ($\log g$) for each star were determined 
from colors of Str\"{o}mgren's $uvby\beta$ photometric system 
by using Napiwotzki et al.'s (1993) calibration. 
 
Especially important parameter we should care about is the microturbulence 
($\xi$). We basically adopted (as done in T18) the analytical 
$T_{\rm eff}$-dependent relation derived by Takeda et al. (2008)
\begin{equation}
\xi = 4.0 \exp\{- [\log (T_{\rm eff}/8000)/A]^{2}\} 
\end{equation}
(where $A \equiv [\log (10000/8000)]/\sqrt{\ln 2}$, 
$\xi$ is in km~s$^{-1}$, and $T_{\rm eff}$ is in K) for stars with
$T_{\rm eff} < 11000$~K, while $\xi = 1$~km~s$^{-1}$ was assumed
at $T_{\rm eff} > 11000$~K (where this equation yields $\xi < 1$~km~s$^{-1}$). 
Such formula-based values are called as the ``standard'' microturbulence 
(designated as $\xi_{\rm std}$) in order to clarify the difference from 
another choice of microturbulence described later (cf. Section~6.2).

The only exception is the standard star Procyon (HD~61421),\footnote{
The reason why Procyon was chosen as the reference standard (as done 
in our previous studies) is to carry out abundance determination by 
``differential analysis'' where the resulting relative abundances are 
unaffected by uncertainties in the $gf$ values of spectral lines. That is, 
Procyon (F5 IV--V) is more suitable than the Sun (whose $T_{\rm eff}$ is 
too low in comparison with those of A and late B stars to be used for 
such a purpose), while its chemical abundances are practically the same 
as those of the Sun (cf. the references quoted in Section IV(c) of 
Takeda et al. 2008).}
for which we used Takeda et al.'s (2005b) spectroscopically determined values
($T_{\rm eff}$ = 6612~K, $\log g = 4.00$, and $\xi_{\rm std} = 1.97$~km~s$^{-1}$)
to maintain consistency with Takeda et al. (2008).

The adopted values of $T_{\rm eff}$, $\log g$, [Fe/H],\footnote{
These Fe abundances were already established in our previous papers 
(cf. the references given in Table 2) based on the spectrum-fitting 
method in the wavelength region ($\sim$~20--30\AA\ wide) centered 
around $\sim$~6155\AA\ (where the Fe~{\sc ii} 6147/6149 doublet lines
are the important indicators of Fe abundance).}
and $\xi_{\rm std}$ are summarized in Table~1.
All the program stars are plotted on the $\log L$ vs. $\log T_{\rm eff}$ 
diagram (theoretical HR diagram) in Fig.~1, where theoretical evolutionary 
tracks corresponding to different stellar masses are also depicted. This figure 
indicates that the masses of our sample stars are in the range between 
$\sim 1.5 M_{\odot}$ and $\sim 5 M_{\odot}$. More detailed data regarding the 
targets and their stellar parameters are given in the electronic table (tableE.dat).

The model atmosphere corresponding to each star was constructed by interpolating 
Kurucz's (1993a) ATLAS9 model grid (for $\xi = 2$~km~s$^{-1}$) in terms of 
$T_{\rm eff}$, $\log g$, and [Fe/H]. 

\section{Non-LTE calculation for Si}

The statistical-equilibrium calculations for silicon atom were carried out 
by using the non-LTE code described in Takeda (1991). 
The atomic model of Si adopted in this study was constructed 
based on Kurucz and Bell's (1995) compilation of atomic data ($gf$ values, 
levels, etc.), which consists of 34 Si~{\sc i} terms (up to 4$d$~$^{1}$F$^{\rm o}$
at 58893.4~cm$^{-1}$) with 222 Si~{\sc i} radiative transitions, 
31 Si~{\sc ii} terms (up to 3$p^{3}$~$^{4}$S$^{\rm o}$
at 123033.5~cm$^{-1}$) with 109 Si~{\sc ii} radiative transitions, 
and 23 Si~{\sc iii} terms (up to 4$p$~$^{3}$P at 248073~cm$^{-1}$;
included only for conservation of total Si atoms). 

Regarding evaluations of photoionization rates, the cross-section data taken 
from TOPbase (Cunto, Mendoza 1992) were used for the lower 10 Si~{\sc i} terms and 
10 Si~{\sc ii} terms (while hydrogenic approximation was assumed for higher terms). 
As to the collisional rates, the theoretical results of Aggarwal and Keenan (2014) 
were invoked for the bound--bound electron impact rates between the 
lower 10 Si~{\sc ii} terms. Otherwise, the recipe described in Sect.~3.1.3 
of Takeda (1991) was followed (inelastic collisions due to neutral hydrogen 
atoms were formally included as described therein, though insignificant 
in the atmosphere of early-type stars under question).   

The calculations were done on a grid of 44 ($= 11 \times 4$) 
solar-metallicity ([Fe/H] = 0) model atmospheres 
resulting from combinations of eleven $T_{\rm eff}$ values 
(6500, 7000, 7500, 8000, 8500, 9000, 9500, 10000, 11000, 12000, 13000, 
and 14000~K) and four  $\log g$ values (3.0, 3.5, 4.0, and 4.5) while 
assuming $\xi$ = 2~km~s$^{-1}$ and the Si abundance of $A$(Si) = 7.55 
(solar Si abundance adopted in ATLAS9 models). 
The depth-dependent non-LTE departure coefficients to be used for each star were 
then evaluated by interpolating this grid in terms of $T_{\rm eff}$ and $\log g$.

\section{Abundance determination}

The non-LTE Si abundances were determined (as done in T18 for CNO abundances) 
based on Takeda's (1995) numerical algorithm by accomplishing the best fit between 
the synthetic and observed spectra in the 6340--6380~\AA\ region while varying 
the abundances of Si and some other elements showing appreciable lines (especially 
Fe, plus other elements such as Mg, Mn, Zn depending on cases), $v_{\rm M}$ 
(macrobroadening velocity corresponding to instrumental/rotational broadening 
and macroturbulence) and $\Delta \lambda$ (radial velocity or wavelength shift) 
but the microturbulence being fixed at $\xi_{\rm std}$. 
Since the relevant wavelength region of the raw spectra is more or less contaminated
by weak telluric lines, they were removed in advance by dividing by the spectrum 
of a rapid rotator as demonstrated in Fig.~2.
The atomic data of spectral lines comprising in this region were exclusively taken 
from Kurucz and Bell's (1995) compilation (those of relevant Si~{\sc ii} doublet
lines are summarized in Table~3), though some pre-adjustments\footnote{
Six lines (Ca~{\sc i} 6343.308, Si~{\sc i} 6353.360, Fe~{\sc i} 6353.835,
Si~{\sc i} 6356.321, Ca~{\sc i} 6361.786, and Fe~{\sc i} 6368.620) included
in this database were neglected, while the $\log gf$ of Fe~{\sc i} 6358.631 
was changed from $-1.04$ to $-1.70$.} were necessary in order to achieve an satisfactory fit.
The accomplished fit in the neighborhood of both lines for each star is displayed 
in Fig.~3.

\begin{figure}
\centerline{\includegraphics[width=0.7\textwidth,clip=]{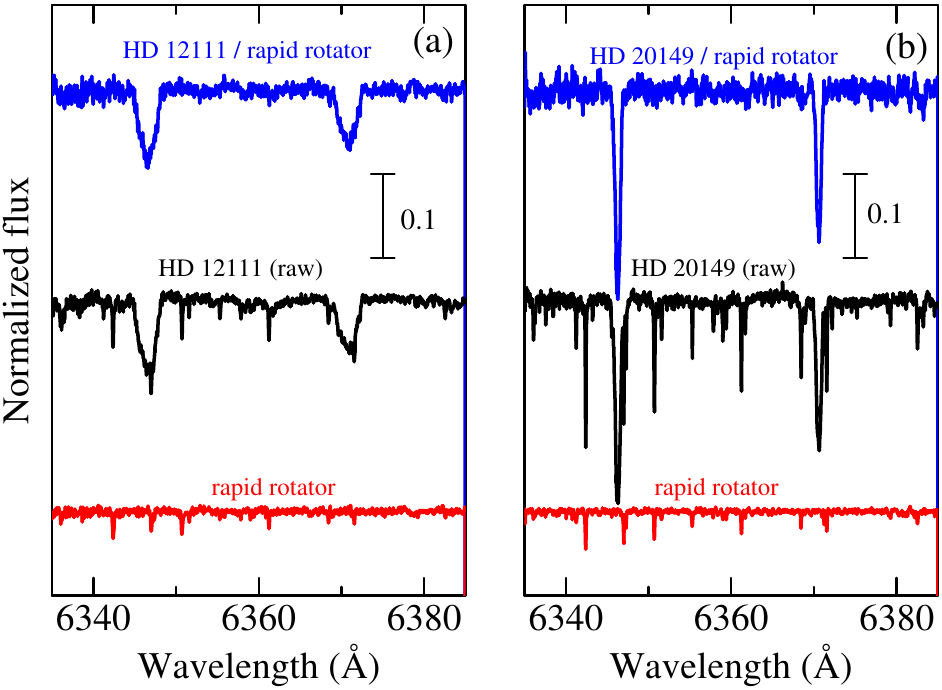}}
\caption{
Examples of how the telluric lines (mostly due to
H$_{2}$O vapor) are removed in the 6340--6380~\AA\ region
comprising Si~{\sc ii} 6347/6371 lines. 
Dividing the raw stellar spectrum (middle, black) by the spectrum 
of a rapid rotator (bottom, red) results in the final 
spectrum (top, blue). The left (a) and right (b) panel 
show the cases of HD~12111 (weaker telluric contamination) and HD~20149 
(stronger contamination), respectively. No Doppler correction is applied 
to the wavelength scale of these spectra. 
}
\label{fig2}
\end{figure}

\begin{figure}
\centerline{\includegraphics[width=1.\textwidth,clip=]{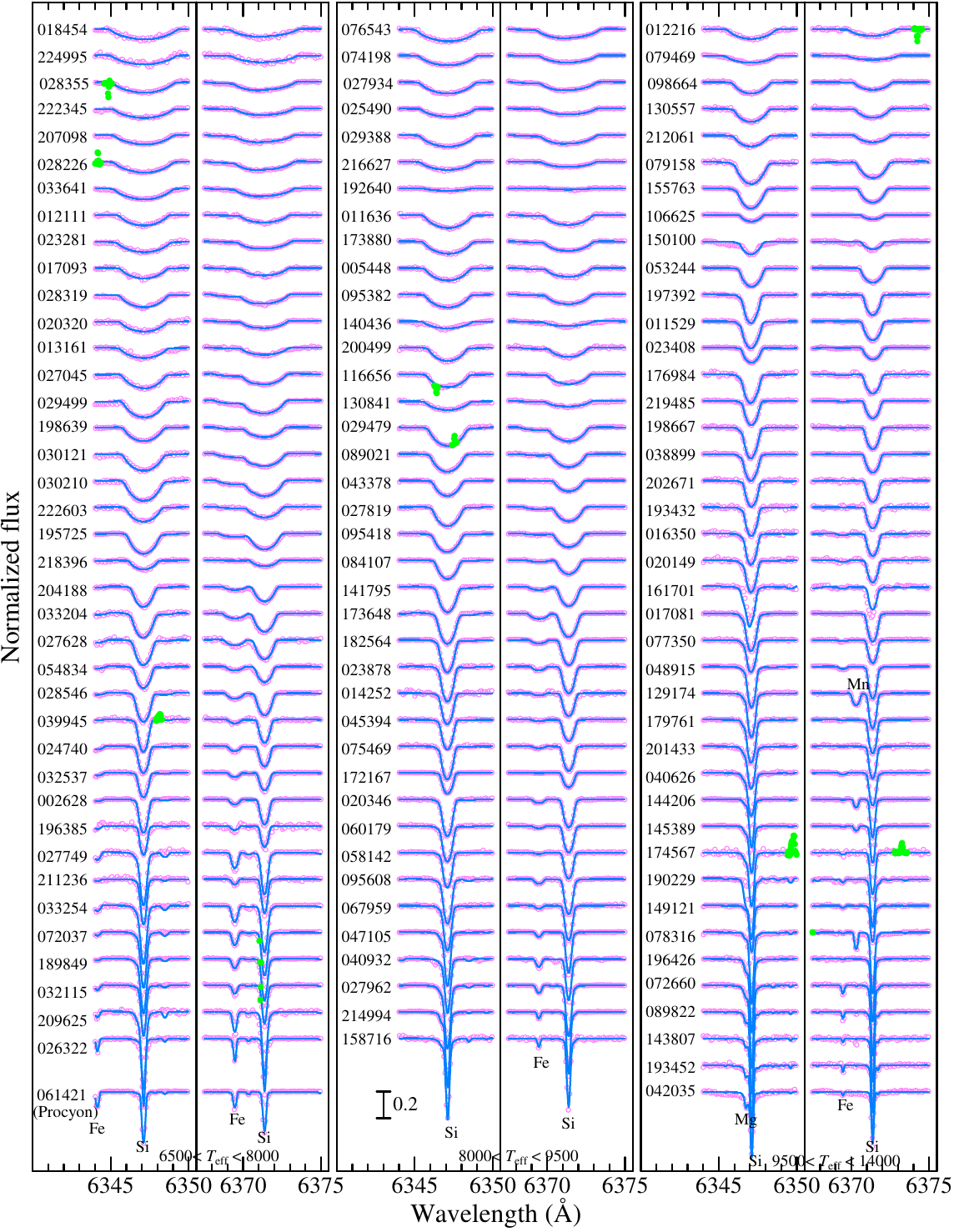}}
\caption{
Synthetic spectrum fitting analysis for Si abundance determination
from Si~{\sc ii} 6347/6371 lines,
In the left, middle, and right panels are shown the results 
for 40 stars in 6500~K~$< T_{\rm eff} < 8000$~K,
39 stars in 8000~K~$< T_{\rm eff} < 9500$~K, and
41 stars in 9500~K~$< T_{\rm eff} < 14000$~K, respectively.
The best-fit theoretical spectra (in the selected ranges of 
6344--6350~\AA\ and 6367.5--6375~\AA\ comprising the relevant 
Si~{\sc ii} lines) are depicted by blue solid lines, 
while the observed data are plotted by pink symbols (the masked data 
excluded in judging the goodness of fit are highlighted in green).  
In each panel, the spectra are arranged in the descending order 
of $v_{\rm e} \sin i$, and an offset of 0.2 is 
applied to each spectrum (indicated by the HD number) relative to 
the adjacent one. The case of Procyon (standard star) is 
separately displayed at the bottom of the left panel.
}
\label{fig3}
\end{figure}

\setcounter{table}{2}
\begin{table}
\small
\caption{Adopted atomic data of Si~{\sc ii} 6347 and 6371 lines.}
\begin{center}
\begin{tabular}
{crrrrrr}\hline \hline
Multiplet & $\lambda$ & $\chi_{\rm low}$ & $\log gf$ & Gammar & Gammas & Gammaw \\
  No.  &  (\AA)  & (eV)  &  (dex)  & (dex)  & (dex)  &  (dex) \\
\hline
2  & 6347.109 & 8.121 &  +0.297  & 9.09 & $-5.04$ & $(-7.68)$ \\
2  & 6371.371 & 8.121 & $-0.003$ & 9.08 & $-5.04$ & $(-7.68)$ \\
\hline
\end{tabular}
\end{center}
\scriptsize
Note. \\
These data are were taken from Kurucz and Bell's (1995) compilation,
while those parenthesized are the default values calculated
by Kurucz's (1993a) WIDTH9 program.\\
Followed by first four self-explanatory columns,
damping parameters are given in the last three columns:\\
Gammar is the radiation damping width (s$^{-1}$), $\log\gamma_{\rm rad}$.\\
Gammas is the Stark damping width (s$^{-1}$) per electron density (cm$^{-3}$) 
at $10^{4}$ K, $\log(\gamma_{\rm e}/N_{\rm e})$.\\
Gammaw is the van der Waals damping width (s$^{-1}$) per hydrogen density 
(cm$^{-3}$) at $10^{4}$ K, $\log(\gamma_{\rm w}/N_{\rm H})$. \\
\end{table}

Then, the equivalent widths ($W_{6347}$ and $W_{6371}$) of the Si~{\sc ii} 6347 
and 6371 lines were inversely evaluated from the best-fit solution of 
$A^{\rm N}_{\rm std}$(Si) with the same model and atmospheric parameters as 
used in the spectrum-fitting analysis. 
From such evaluated $W$, the non-LTE abundance ($A^{\rm N}$), LTE abundance ($A^{\rm L}$) 
and non-LTE correction ($\Delta \equiv A^{\rm N} - A^{\rm L}$) were derived for each line. 
Besides, $W$ can be further used to estimate the abundance uncertainties 
due to typical ambiguities of atmospheric parameters by perturbing the standard 
values interchangeably. 
Such derived $W_{6347}$/$W_{6371}$ (equivalent widths), $\Delta_{6347}$/$\Delta_{6371}$ 
(non-LTE corrections), $A^{\rm N}_{\rm std}$ (non-LTE Si abundance), $\delta_{T\pm}$ 
(abundance changes for $T_{\rm eff}$ perturbations by $\pm 3$\%), $\delta_{g\pm}$ 
(abundance changes for $\log g$ perturbations by $\pm 0.1$~dex), and $\delta_{\xi\pm}$ 
(abundance changes for $\xi$ perturbations by $\pm 30$\%) are plotted against 
$T_{\rm eff}$ in Fig.~4.

These standard abundances, expressed in the form of [Si/H]$_{\rm std}$ 
($\equiv A^{\rm N}_{\rm std}$(star) $-$ $A^{\rm N}_{\rm std}$(Procyon)) 
are given in Table~1. More complete results including $W$ and $\Delta$ are 
presented in ``tableE.dat'' of the online material.

\begin{figure}
\centerline{\includegraphics[width=0.7\textwidth,clip=]{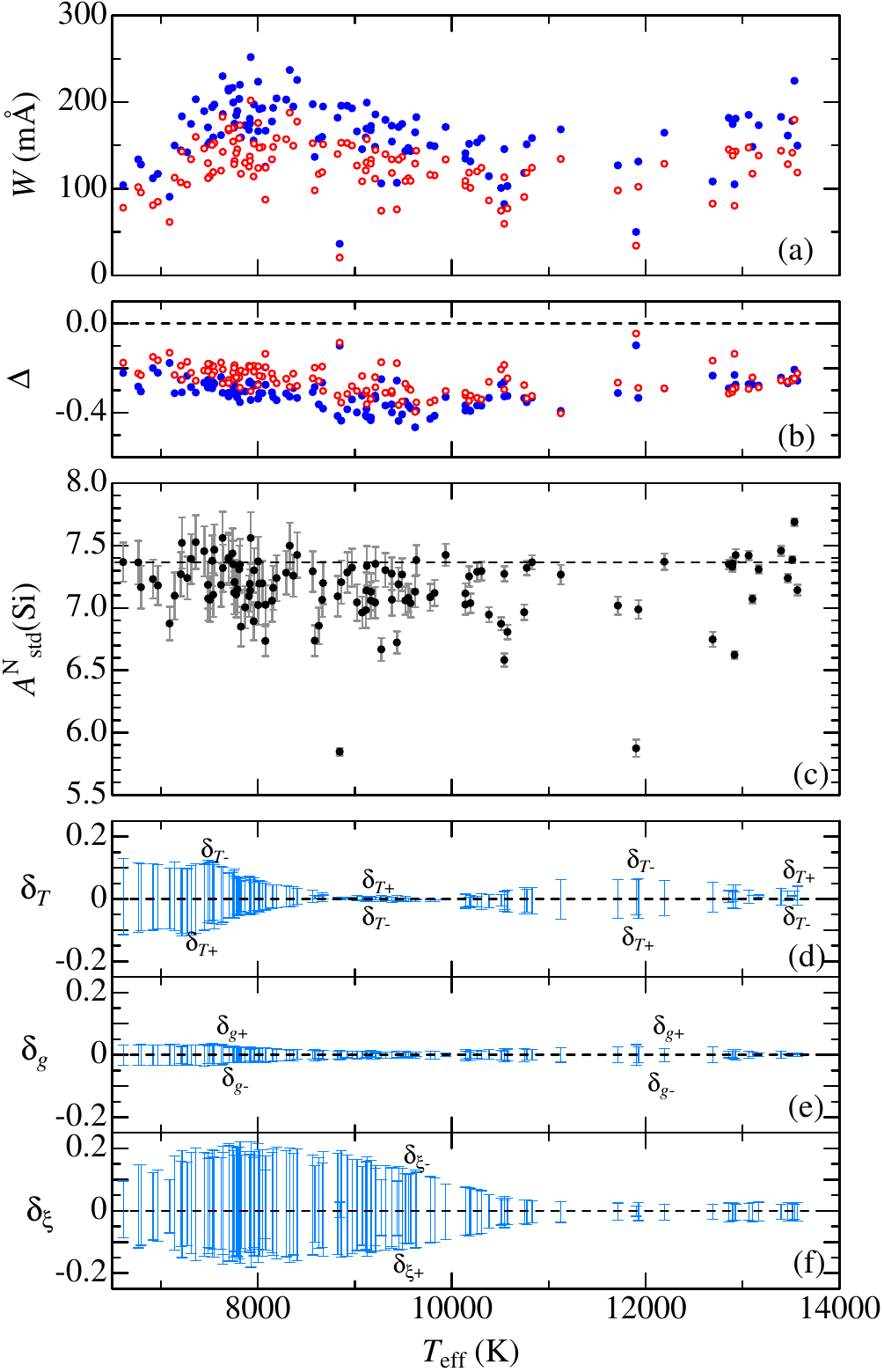}}
\caption{
Silicon abundances and the related quantities plotted against $T_{\rm eff}$. 
(a) Equivalent widths of Si~{\sc ii} 6347 ($W_{6347}$, filled symbols) and 
Si~{\sc ii} 6371 ($W_{6371}$, open symbols). 
(b) Non-LTE corrections for  Si~{\sc ii} 6347 ($\Delta_{6347}$, filled symbols) 
and Si~{\sc ii} 6371 ($\Delta_{6371}$, open symbols)
(c) $A_{\rm std}^{\rm N}$(Si) (standard non-LTE Si abundance corresponding 
to $\xi_{\rm std}$), where the error bar denotes $\pm\delta_{Tgv}$ 
(root-sum-square of $\delta_{T}$, $\delta_{g}$, and $\delta_{\xi}$, 
where $\delta_{T}$ is the mean of $|\delta_{T+}|$ and $|\delta_{T-}|$; etc.),
(d) $\delta_{T+}$ and $\delta_{T-}$ (abundance variations for Si~{\sc ii} 6347
in response to $T_{\rm eff}$ changes of +3\% and $-$3\%), 
(e) $\delta_{g+}$ and $\delta_{g-}$ (abundance variations for Si~{\sc ii} 6347
in response to $\log g$ changes by $+0.1$~dex and $-0.1$~dex), 
and (f) $\delta_{\xi +}$ and $\delta_{\xi -}$ (abundance variations for 
Si~{\sc ii} 6347 in response to perturbing the $\xi_{\rm std}$ 
value by +30\% and $-$30\%).
The abundance of Procyon ($A^{\rm N}_{\rm std} = 7.367$), which is adopted 
as the reference, is indicated by the horizontal dashed line in panel (c). 
}
\label{fig4}
\end{figure}

\section{Discussion and conclusion}

\subsection{Characteristics of the non-LTE effect}

As seen from the results derived in Section~5, the Si~{\sc ii} 6347/6371 lines
suffer an appreciable non-LTE effect. According to Fig.~4b, their 
non-LTE abundance corrections ($\Delta$) are negative (which means that the 
non-LTE effect strengthens the lines) and typically a few tenths dex 
($|\Delta_{6347}|\sim$~0.2--0.5~dex, $|\Delta_{6371}|\sim$~0.1--0.4~dex; 
naturally the former is larger because of the stronger line forming in 
comparatively shallower layer).  The maximum of $|\Delta|$ is 
around $T_{\rm eff} \sim 10000$~K.  

In Fig.~5 are shown the $l_{0}^{\rm NLTE}(\tau)/l_{0}^{\rm LTE}(\tau)$ 
(the non-LTE-to-LTE line-center opacity ratio; almost equal to 
$\simeq b_{1}$) and $S_{\rm L}(\tau)/B(\tau)$ (the ratio of 
the line source function to the Planck function; nearly equal to 
$\simeq b_{2}/b_{1}$) for the transition relevant to the Si~{\sc ii} 6347/6371 lines
($b_{1}$ and $b_{2}$ are the non-LTE departure coefficients for the lower and
upper terms), which were computed on the models of representative $T_{\rm eff}$ 
and $\log g$ values.  
As seen from this figure, while $l_{0}^{\rm NLTE}/l_{0}^{\rm LTE} > 1$ 
(overpopulation) holds in the line-forming region at $T_{\rm eff} \la 10000$~K
(A-type tars), this inequality suddenly turns to be reversed (underpopulation) 
at higher $T_{\rm eff}$ (late B-type stars) because of the beginning of 
Si~{\sc ii} overionization (once-ionized Si is not the dominant ionization 
stage any more in such a higher $T_{\rm eff}$ regime). Although the non-LTE effect still acts 
to intensify lines ($\Delta$ remains negative) at 10000~K~$\la T_{\rm eff} \la 14000$~K 
due to the dilution of $S_{\rm L} (< B)$ (see the lower panels in Fig.~5), 
$|\Delta|$ progressively decreases with an increase in $T_{\rm eff}$ (see also
the Appendix~A where the behavior of $\Delta$ in B-type stars is further discussed).

How the theoretical $W$ and $\Delta$ computed for these two Si~{\sc ii} lines 
depend upon the atmospheric parameters ($T_{\rm eff}$, $\log g$, and $\xi$) 
is illustrated in Fig.~6, which reasonably explains the trends observed 
in Figs.~4a and 4b (the maximum of $W$ is seen around $T_{\rm eff} \sim 8000$~K 
because the peak of $\xi$ is attained there). 

\begin{figure}
\centerline{\includegraphics[width=0.8\textwidth,clip=]{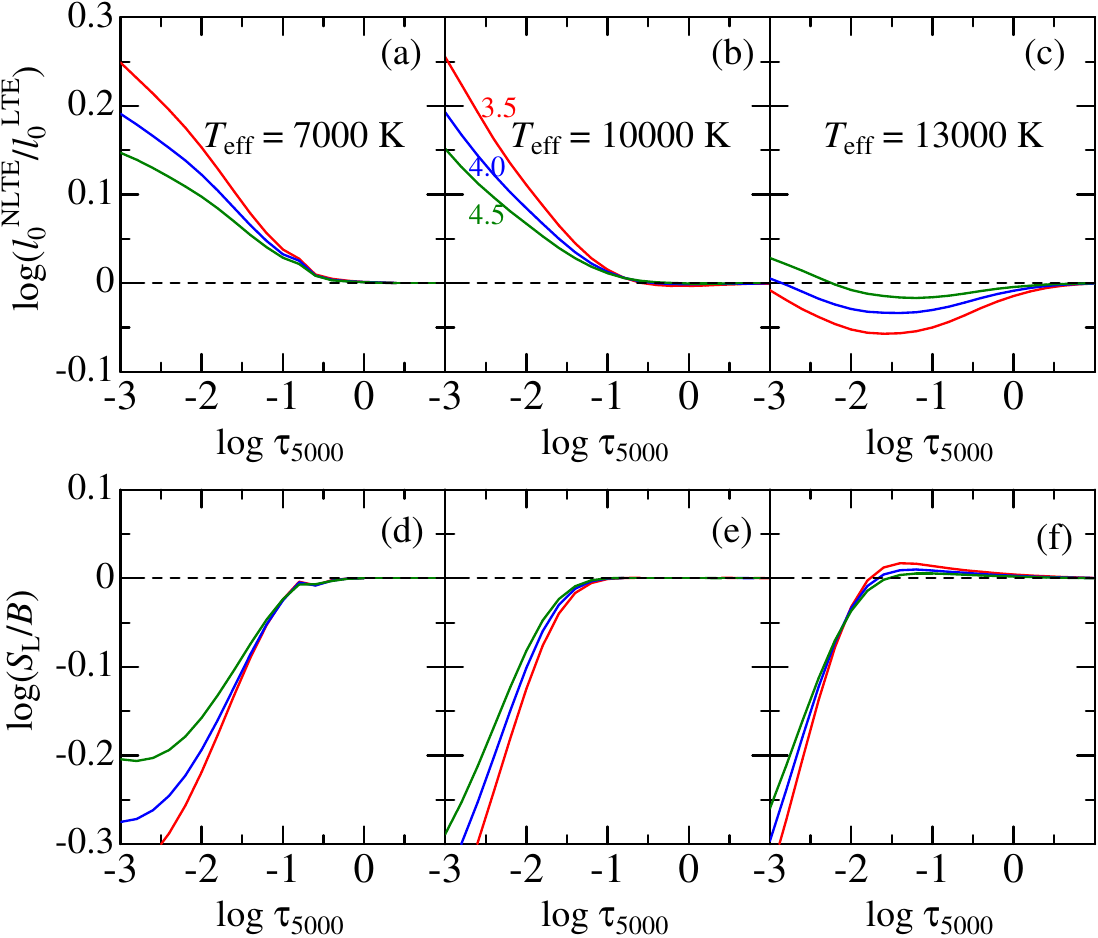}}
\caption{
The non-LTE-to-LTE line-center opacity ratio (upper panels a--c) and 
the ratio of the line source function ($S_{\rm L}$) 
to the local Planck function ($B$) (lower panels d--f)  
for the Si~{\sc ii} 4s~$^{2}$S--4p~$^{2}$P$^{\rm o}$ transition 
(corresponding to Si~{\sc ii} 6347/6371 lines) of multiplet~2, 
plotted against the continuum optical depth at 5000~\AA. 
Computations were done with $\xi = 2$~km~s$^{-1}$ on the solar-metallicity 
models ([Fe/H] = [Si/Fe] = 0) of $T_{\rm eff} =$ 7000~K (left panels a, d), 
10000~K (middle panels b, e), and 13000~K (right panels c, f).
At each panel are shown the results for three $\log g$ values of 3.5, 
4.0, and 4.5 depicted by different colors (red, blue, and green, respectively). 
}
\label{fig5}
\end{figure}

\begin{figure}
\centerline{\includegraphics[width=0.8\textwidth,clip=]{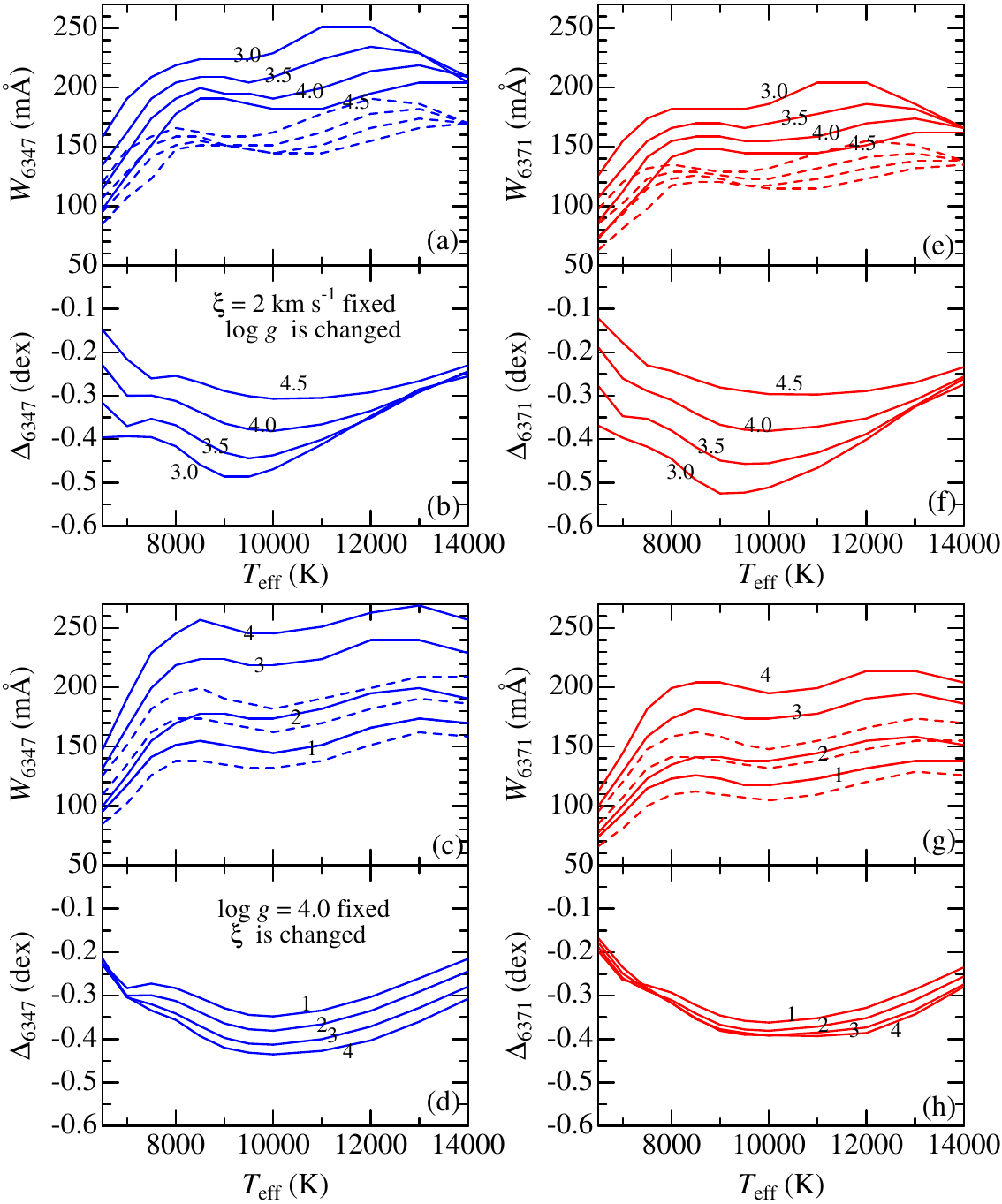}}
\caption{
The non-LTE and LTE equivalent widths ($W^{\rm N}$ and $W^{\rm L}$) 
for the Si~{\sc ii} 6347/6371 lines and the corresponding non-LTE corrections 
($\Delta$), which were computed on the non-LTE grid of models described in 
Section 4, are plotted against $T_{\rm eff}$.
Each figure set consists of two panels; the upper panel is for $W^{\rm N}$ 
(solid lines) and $W^{\rm L}$ (dashed lines), while the lower panel is
for $\Delta$. The upper sets (a+b, e+f) show the case of fixed $\xi$ 
(2~km~s$^{-1}$) but different $\log g$ (3.0, 3.5, 4.0, and 4.5), 
while the lower sets (c+d, g+h) are for the case of fixed $\log g$ (4.0) 
but different $\xi$ (1, 2, 3, and 4~km~s$^{-1}$). The left-hand figures 
show the results for the Si~{\sc ii} 6347 line, while the right-hand
ones for the Si~{\sc ii} 6371 line.
}
\label{fig6}
\end{figure}

\subsection{Consistency check of microturbulence}

As is evident from the lower three panels (d--f) of Fig.~4, an uncertainty $\xi$
has the most significant impact on the Si abundance among the three atmospheric 
parameters especially at $T_{\rm eff} \la 10000$~K, since the Si~{\sc ii} 6347/6371 lines
are strong and saturated (on the flat part of the curve of growth).
Therefore, particular attention should be paid to whether or not an appropriate 
choice of $\xi$ has be done. As a matter of fact, Takeda et al. (2009) reported that
considerably underestimated Na abundance would result from the strongly saturated 
Na~{\sc i} 5889/5895 D lines if the microturbulence given by Equation~(1) is used,
which may be attributed to the depth-dependence of $\xi$ (cf. Section~5 therein)
Does such an inadequacy similarly exist also for the case of Si~{\sc ii} 6347/6371 lines?

In order to examine this problem, another solution of microturbulence was determined 
from these doublet lines themselves by taking the advantage that their $\log gf$ 
strengths are different by 0.3~dex. That is, spectrum fitting analysis was 
retried (taking $A^{\rm N}_{\rm std}$ and $\xi_{\rm std}$ as the starting solutions) 
while allowing {\it both} $A^{\rm N}$(Si) and $\xi$ to vary. 
These refit solutions (which are referred to as $A^{\rm N}_{\rm refit}$ and 
$\xi_{\rm refit}$) were successfully converged for 97 stars (about $\sim 80$\%), 
though failed for the remaining 23 stars. 

The resulting $A^{\rm N}_{\rm refit}$ and $\xi_{\rm refit}$ (given in Table~1) are 
compared with $A^{\rm N}_{\rm std}$ and $\xi_{\rm std}$ in Fig.~7, where 
the following characteristics are observed.\\
--- Fig.~7a indicates that consistency between $\xi_{\rm refit}$ (dots) and
$\xi_{\rm std}$ (solid line) is not necessarily bad, though considerable
discrepancy (quite a few $\xi_{\rm refit}$ values tending to be appreciably 
lower than $\xi_{\rm std}$) is seen around $T_{\rm eff} \sim 8000$~K.\footnote{
Besides, two $\xi_{\rm refit}$ values around $T_{\rm eff} \sim$~13000~K
are apparently two large; but they are not reliable and should not be seriously 
taken, because determination becomes more difficult for the case of weaker 
Si~{\sc ii} lines at higher $T_{\rm eff}$.}\\
--- As a result, $A_{\rm std}^{\rm N}$ tends to be lower than $A_{\rm refit}^{\rm N}$ 
at $T_{\rm eff} \sim 8000$~K. The differences are typically a few tenths dex  
(four stars show especially large discrepancies of $\sim$~0.5--0.6~dex (cf. Fig.~7c).\\ 
--- It is worth noting that the [Si/H]$_{\rm std}^{\rm N}$ values also exhibit 
similar discrepancies when compared with Takeda et al.'s (2009) [Si/H] results derived 
from the spectrum fitting in the 6140--6170~\AA\ region comprising Si~{\sc i} lines (Fig.~7d).\\
--- Accordingly, we may state that the abundances derived from Si~{\sc ii} 6347/6371
lines by using Equation~(1)-based $\xi_{\rm std}$ are apt to be underestimated
around $T_{\rm eff} \sim 8000$~K corresponding to late-to-mid A-type stars.\\
--- However, this problem is not so serious as the case of Na~{\sc i} 5889/5895 
lines addressed by Takeda et al. (2009). Actually, the appearance of 
$A_{\rm refit}^{\rm N}$ vs. $T_{\rm eff}$ plot (Fig.~7b) is not 
significantly different from the case of $A_{\rm std}^{\rm N}$ (Fig.~4c).
In the figures illustrating the behaviors of Si abundances to be discussed 
in the next section, both (``std'' and ``refit'') results will be shown,
so that they may be compared with each other.  

\begin{figure}
\centerline{\includegraphics[width=0.8\textwidth,clip=]{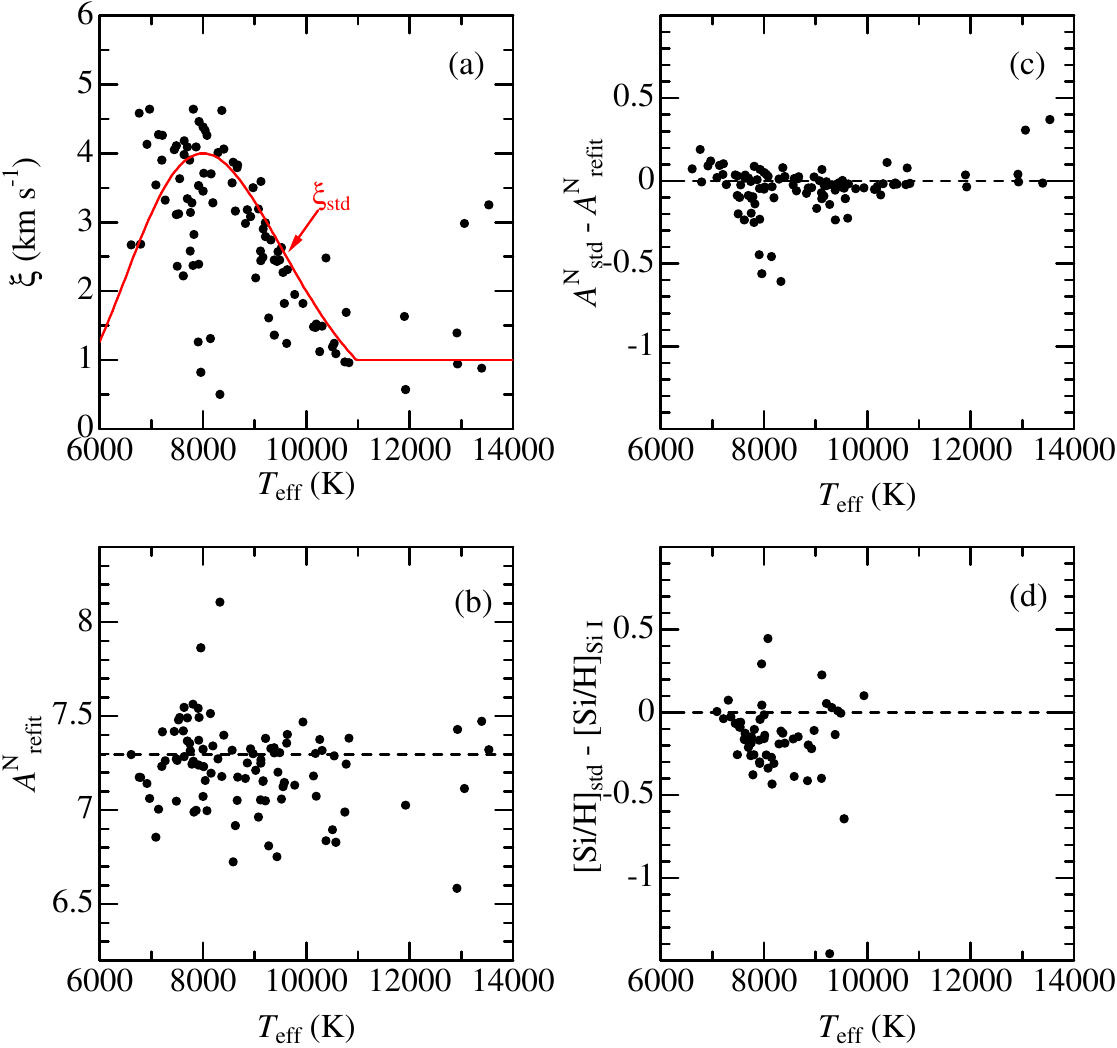}}
\caption{
(a) Microturbulence directly determined by spectrum refitting 
($\xi_{\rm refit}$) plotted against $T_{\rm eff}$ by dots, while the 
$T_{\rm eff}$-dependent standard microturbulence ($\xi_{\rm std}$) 
given by Equation (1) is shown by the red solid line.
(b) Non-LTE Si abundance ($A^{\rm N}_{\rm refit}$) resulting from refitting 
(corresponding to $\xi_{\rm refit}$) plotted against $T_{\rm eff}$.
(c) Difference between ``std'' and ``refit'' abundances 
($A^{\rm N}_{\rm std} - A^{\rm N}_{\rm refit}$) plotted against $T_{\rm eff}$.
(d) Difference between [Si/H]$_{\rm std}$ (derived in this study based on 
Si~{\sc ii} 6347/6371 lines by using the standard $\xi_{\rm std}$) and
[Si/H]$_{\rm Si~I}$ (derived by Takeda et al. 2009) based on the spectrum 
fitting applied to the region comprising Si~{\sc i} lines) plotted 
against $T_{\rm eff}$.   
}
\label{fig7}
\end{figure}

\subsection{Observed trend of Si abundances}

\begin{figure}
\centerline{\includegraphics[width=1.\textwidth,clip=]{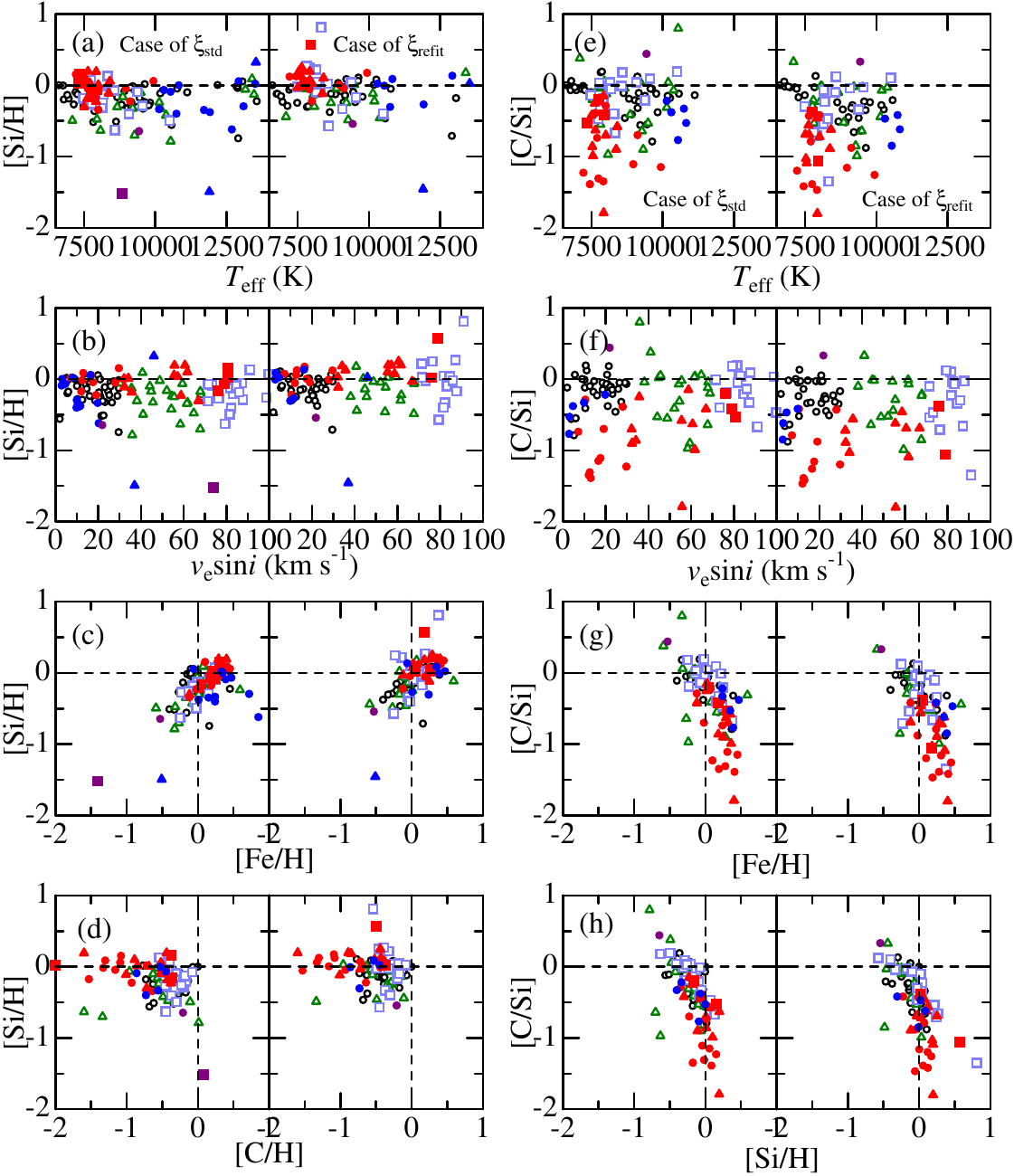}}
\caption{
Graphical display of how [Si/H] and [C/Si] for the 120 program stars are 
related to stellar parameters (or abundances).
Left panels: [Si/H] (Si abundance relative to Procyon) plotted against
(a) $T_{\rm eff}$, (b) $v_{\rm e}\sin i$, (c) [Fe/H], and (d) [C/H].
Right panels: [C/Si] (logarithmic C-to-Si abundance ratio) plotted against
(e) $T_{\rm eff}$, (f) $v_{\rm e}\sin i$, (g) [Fe/H], and (h) [Si/H].
Each panel consists of two similar diagrams constructed from different Si abundances
$A^{\rm N}_{\rm std}$ (left) and $A^{\rm N}_{\rm refit}$ (right) corresponding 
to $\xi_{\rm std}$ and $\xi_{\rm refit}$, respectively. 
Stars of different $v_{\rm e}\sin i$ classes are discriminated by the types 
of symbols: circles ($0 < v_{\rm e}\sin i < 30$~km~s$^{-1}$), 
triangles ($30 \le v_{\rm e}\sin i < 70$~km~s$^{-1}$), 
and squares ($70 \le v_{\rm e}\sin i < 100$~km~s$^{-1}$).
Normal stars are shown by open symbols, while those classified as chemically peculiar 
are highlighted by filled symbols (red-filled symbols for Am stars, blue-filled ones 
for HgMn stars, and purple-filled ones for $\lambda$~Boo stars).
}
\label{fig8}
\end{figure}

\begin{figure}
\centerline{\includegraphics[width=1.\textwidth,clip=]{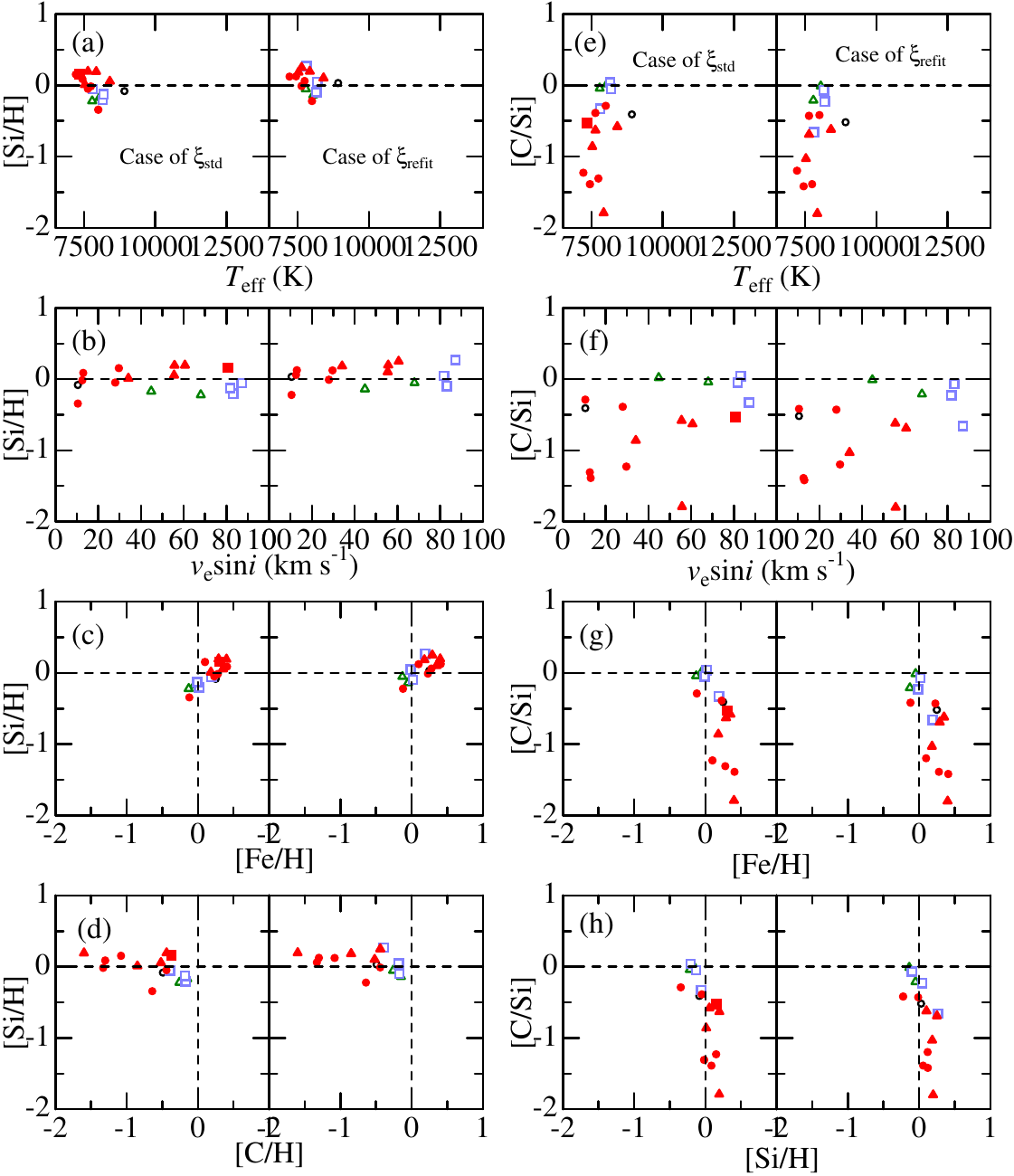}}
\caption{
Graphical display of how [Si/H] and [C/Si] for the 16 Hyades cluster stars 
are related to stellar parameters (or abundances).
Otherwise, the same as in Fig.~8.
}
\label{fig9}
\end{figure}

The relative abundances of Si ([Si/H]) and the C-to-Si ratios ([C/Si] = [C/H]$-$[Si/H]) 
for the 120 stars are plotted against the stellar parameters (and the corresponding 
[C/H] or [Si/H]) in Fig.~8, where two kinds of results based on $A^{\rm N}_{\rm std}$
and $A^{\rm N}_{\rm refit}$ are presented in parallel in each panel. Besides, the same 
correlation plots as Fig.~8 but only for the selected 16 Hyades stars are depicted 
in Fig.~9. The following characteristics can be read from these figures.
\renewcommand{\labelitemi}{$\bullet$}
\begin{itemize}
\item
The resulting Si abundances (relative to Procyon) for most stars are in 
the range of $-0.5 \la$~[Si/H]~$\la +0.3$ (tending to be rather Si-deficient 
than Si-rich).
\item
As for the relation to stellar parameters, any clear dependence upon $T_{\rm eff}$ 
or $v_{\rm e}\sin i$ is not observed in [Si/H] (Figs.~9a and 9b).
\item
Am stars and HgMn stars appear to show somewhat higher [Si/H] than normal stars,
while $\lambda$~Boo stars are naturally Si-deficient. 
The mean [Si/H]$_{\rm std}$ values for each star group are $-0.23$ (normal), 
$-0.03$ (Am), $-0.18$ (HgMn),\footnote{
HD~106625 ($\gamma$ Crv) was excluded in the averaging process because
of its exceptionally low [Si/H]$_{\rm std}$ of $-1.49$ for this HgMn group.} 
 and $-1.08$ ($\lambda$~Boo).
\item
A positive correlation exists between [Si/H] and [Fe/H] (Fig.~8c),
which is also observed for the selected sample of Hyades stars (Fig.~9c).
Actually, the correlation coefficients calculated between [Si/H]$_{\rm std}$ 
and [Fe/H] are  +0.65 (all sample) and +0.82 (Hyades sample).
\item
The C-to-Si ratio ([C/Si]) tends to decrease systematically with an increase 
in [Si/H] (Fig.~8h) indicating that C and Si are anti-correlated, though the 
nature of anti-correlation between [Si/H] and [C/H] is not very clear (Fig.~8d).  
\end{itemize}

How can we interprete these results? As mentioned in Section~1, two physical
processes may be considered for the possible cause of chemical abundance anomalies:
(i) atomic diffusion and (ii) gas--dust separation. 
Regarding the former diffusion process, although considerable uncertainties
still exist, available theoretical calculations predict a deficiency of C, 
a slight underabundance of Si, and an overabundance of Fe (cf. Richer et al. 2000;
Talon et al. 2006). As to the latter gas--dust separation process, Si as well as 
Fe (both are refractory elements and should behave similarly) are expected to 
be anti-correlated with C (volatile species).

Although [C/Si] apparently decreases with an increase in [Si/H], this trend can not 
be simply explained by the gas--dust separation alone, because the (negative) gradient 
of $d$[C/Si]/$d$[Si/H] in Fig.~8h is too steep to be identified with Holweger and 
St\"{u}renburg's (1993) Fig.~2 (the slope is $\sim -45^{\circ}$). Actually, the 
considerably wide range of [C/Si] ($\sim 2$~dex) is mainly due to the diversified
deficiency of C ($-1.5 \la$~[C/H]~$\la 0$ especially seen in Am stars of lower 
$T_{\rm eff}$; cf. Fig.~8d) while the contribution of [Si/H] is comparatively minor.
Therefore, the main cause of C deficiency is attributed to (not the dust--gas separation 
but) to atomic diffusion as discussed in T18.

Regarding the cause for the dispersion of [Si/H], the fact that [Si/H] and [Fe/H] 
correlate well with each other (Fig.~8c, Fig.~9c; just like Holweger and 
St\"{u}renburg's Fig.~3) may indicate that the gas--dust separation process 
(acting both Si and Fe in a same direction) is involved at least partly, because 
atomic diffusion would differently affect Si and Fe  according to the currently 
available calculations. However, it is unlikely that the diffusion process does 
not play any role in affecting the surface abundances of Si and Fe, since Takeda 
and Sadakane (1997) reported the $v_{\rm e}\sin i$-dependence of 
over-/under-abundance of Fe/O in Hyades A-type stars (cf. Fig.~7 therein)
which is reasonably interpreted as the result of atomic diffusion being more 
effective for slower rotators. 

Consequently, given the available information alone, it is hardly possible to find 
any satisfactory interpretation regarding the observed behavior of Si abundances 
(particularly in relation to the abundances of C and Fe). It may be possible that 
both processes (atomic diffusion and gas--dust separation) concurrently operate 
in an intricate manner (the former being more significant?).
Unfortunately, the current diffusion calculations appear to still suffer 
considerable uncertainties (e.g., in the choice of parameters concerning 
turbulent mixing or mass loss).\footnote{ 
In this context, Takeda et al. (2012) pointed out that [Na/H] well correlate with
[Fe/H] in A-type stars, which contradicts the prediction from the diffusion
theory (Na is expected to be almost normal or slightly underabundant unlike Fe). 
That situation is quite similar to the present case of [Si/H].}
Further progress in this field is desirably awaited, so that it may shed light 
to this complicated situation. 

\acknowledgements

This research has made use of the SIMBAD database, operated at CDS, Strasbourg, France.

\clearpage
\section*{Appendix. Non-LTE effects on Si~{\sc ii} 6347/6371 lines in B-type stars}

Recently, Mashonkina (2020a) carried out an extensive study on the non-LTE line 
formation for silicon (Si~{\sc i}, Si~{\sc ii}, and Si~{\sc iii}) in  main-sequence 
stars of A- and B-type covering the $T_{\rm eff}$ range between 7000 and 20000~K.
Mashonkina's calculation includes the Si~{\sc ii} 6347/6371 lines and 
the non-LTE corrections for these lines derived by her for late A through late B stars 
($T_{\rm eff}\sim$~7000--13000~K; negative $\Delta$ with extents of several tenths dex) 
are more or less consistent with the results of this investigation.

However, her calculation failed to explain the formation of these Si~{\sc ii} 
doublet lines in the early B-type star $\iota$~Her ($T_{\rm eff} = 17500$~K), 
because of the {\it positive} non-LTE corrections resulting in unacceptably 
large non-LTE Si abundances ($\Delta_{6347} = +0.60$, $\Delta_{6371} = +0.67$,
$A^{\rm N}_{6347} = 8.38$, $A^{\rm N}_{6371} = 8.27$; cf. Table~4 in her paper).

Although such an early B-type star is outside of the scope of this study,
it is interesting to examine whether similar inconsistency emerges in our
calculations at the higher $T_{\rm eff}$ regime ($>15000$~K).
For this purpose, additional non-LTE calculations were performed 
for the $\log g = 4$ models with extended $T_{\rm eff}$ up to 20000~K.
The resulting runs of $l_{0}^{\rm NLTE}/l_{0}^{\rm LTE}$ and 
$S_{\rm L}/B$ with depth for the Si~{\sc i} 6347/6371 lines (from 
$T_{\rm eff} = 8000$~K through 20000~K) are depicted in Figs.~10a and 10b; 
and Figs.~10c and 10d display how $W_{6371}$ and $\Delta_{6371}$ 
(for the weaker line of the doublet) vary with $T_{\rm eff}$.

Our calculations suggest that the extent of the (negative) non-LTE abundance 
correction ($|\Delta|$) progressively decreases with an increase of $T_{\rm eff}$ 
in the regime of B-type stars ($T_{\rm eff} \ga 10000$~K), until it eventually 
reaches $\Delta \sim 0$ at the critical $T_{\rm eff}$ of $\sim 19000$~K; 
thereafter $\Delta$ turns into positive (cf. Fig.~10d). In other words,
the line is strengthened by the non-LTE effect ($W^{\rm N} > W^{\rm L}$) 
at $T_{\rm eff} \la 19000$~K while weakened ($W^{\rm N} < W^{\rm L}$) 
at $T_{\rm eff} \ga 19000$~K, as can be confirmed in Fig.~10c.

As mentioned in Section~6.1, the behavior of $\Delta$ is mainly
controlled by the line source function; that is, as long as the inequality 
$\langle S_{\rm L} \rangle < \langle B \rangle$ ($S_{\rm L}$ dilution) 
holds in the line-forming region, $\Delta$ remains negative. 
However, according to Fig.~10b, as $T_{\rm eff}$ is ever increased, 
$\langle S_{\rm L} \rangle$ becomes comparable with or even outweighs 
$\langle B \rangle$, which explains why $\Delta$ approaches 
zero or even turns into positive at higher $T_{\rm eff}$ ($\sim 20000$~K).
 
Fig.~10d suggests that the non-LTE correction for $\iota$~Her ($T_{\rm eff} = 17500$~K) 
expected from our calculation is $\Delta_{6371} \sim -0.1$~dex,
Then, since the LTE abundance is $A^{\rm L}_{6371} = 7.60$ (cf. Table 4 in 
Mashonkina 2020a), the non-LTE Si abundance for $\iota$~Her would make 
$A^{\rm N}_{6371} \sim 7.5$, which is almost consistent with the solar abundance.  
  
Mashonkina's (2020a) $\Delta_{6371}$ vs. $T_{\rm eff}$ relation (taken 
from Table~9 of her paper) is also overplotted for comparison in Fig.~10d.
We can see from this figure that the upturn of Mashonkina's $\Delta_{6371}$
is considerably steeper and $\Delta_{6371} \sim 0$ is attained already at 
$T_{\rm eff}\sim$~13000--14000~K, which is in marked contrast to our calculation 
(critical $T_{\rm eff}$ for $\Delta_{6371} \sim 0$ is at $\sim 19000$~K). 
The reason for this discrepancy is not clear. An inspection of the bottom panel 
of Mashonkina's (2020a) Fig.~1 (in comparison with our Fig.~10a) suggests that 
Si~{\sc ii} levels are largely underpopulated (presumably due to more enhanced 
Si~{\sc ii} overionization) in her calculation.   
We suspect that her procedure of evaluating UV photoionizing radiation field 
may have been rather different, for which we used the opacities included in
Kurucz's (1993a) ATLAS9 program along with Kurucz's (1993b) line opacity 
distribution function as described in Section~3.1.2 of Takeda (1991). 

\begin{figure}
\centerline{\includegraphics[width=0.9\textwidth,clip=]{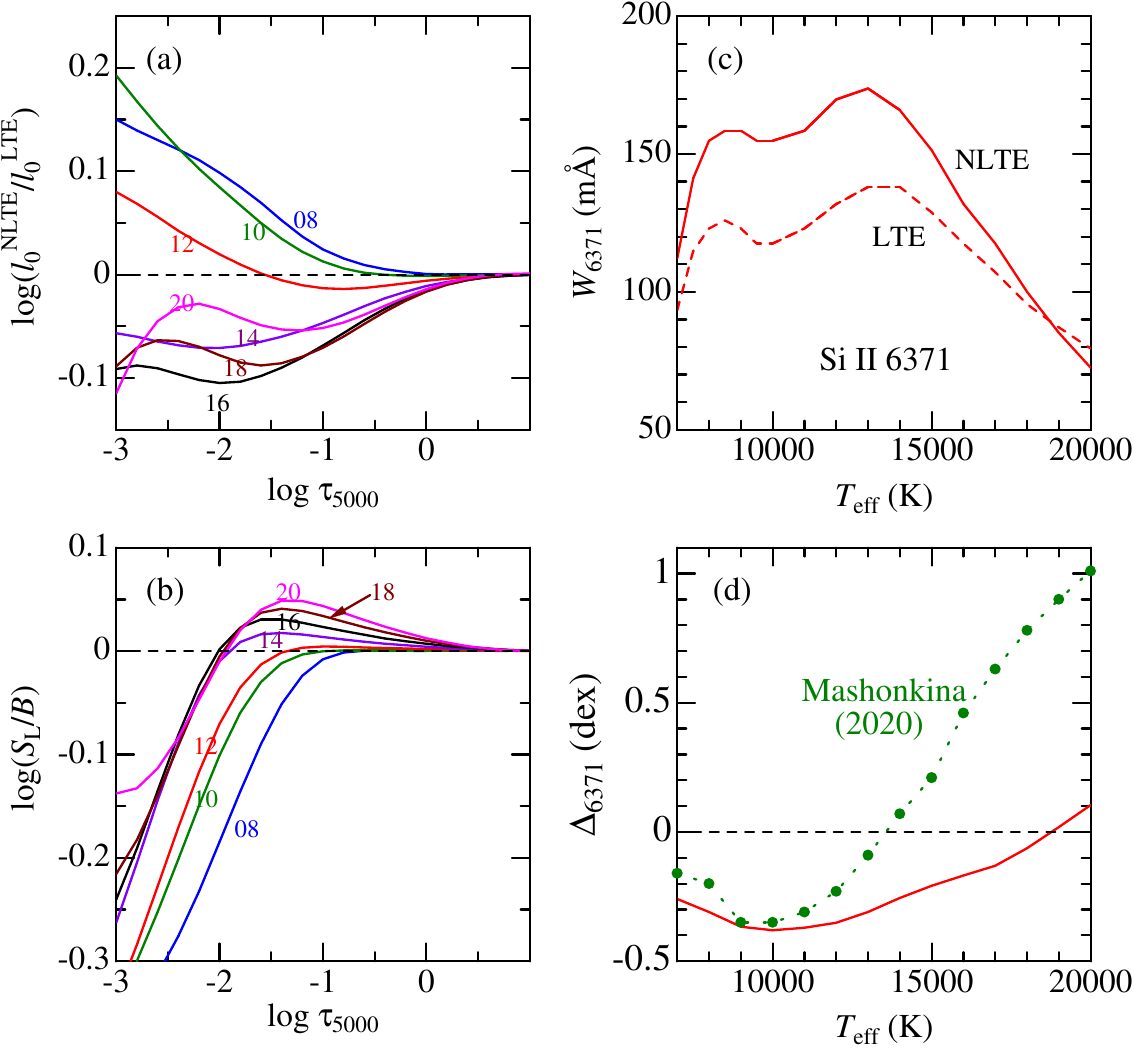}}
\caption{
Non-LTE calculation results for the Si~{\sc ii} 4s~$^{2}$S--4p~$^{2}$P$^{\rm o}$ 
transition of multiplet~2 (like the case of Figs. 5 and 6), which were derived for 
$\log g = 4$ models (with $\xi = 2$~km~s$^{-1}$ and [Si/Fe] = [Fe/H] = 0) 
but $T_{\rm eff}$ extended up to 20000~K (in order to cover early B stars). 
(a) The non-LTE-to-LTE line-center opacity ratio (vs. $\tau_{5000}$),  
(b) $S_{\rm L}/B$ ratio (vs. $\tau_{5000}$), 
(c) non-LTE and LTE equivalent widths for Si~{\sc ii} 6371 (vs. $T_{\rm eff}$),
and (d) non-LTE correction for Si~{\sc ii} 6371 (vs. $T_{\rm eff}$).
In panels (a) and (b), $T_{\rm eff}/1000$ is marked in each curve.
In panel (d), the $T_{\rm eff}$-dependence of $\Delta_{6371}$ calculated
by Mashonkina (2020a) is also shown for comparison. 
}
\label{fig10}
\end{figure}

\clearpage

\begin{center}

\section*{Erratum: \\
Photospheric silicon abundances of 
upper main-sequence stars derived from 
Si~{\sc ii} 6347/6371 doublet lines}
{\it (2024 July 8 by Yoichi Takeda)}
\end{center}

In the article [CAOSP, 52, 5--31 (2022)], Si abundances of 120 late A- 
through late B-type stars were determined by conducting a non-LTE analysis 
on Si~{\sc ii} doublet lines at 6347 and 6371~\AA. It has recently revealed, 
however, that the non-LTE corrections ($\Delta$) and abundances ($A^{\rm N}$) 
derived therein were not correct  because of an inadvertently erroneous treatment 
in the non-LTE calculation program. Specifically, the overionization effect 
of Si~{\sc ii} atoms (acting to weaken Si~{\sc ii} lines or shifting $\Delta$
towards the positive direction) was underestimated by this mistake.
As a consequence, $\Delta$ and $A^{\rm N}$ obtained in that paper were 
more or less undervalued, and this error becomes progressively more 
significant with an increase in $T_{\rm eff}$ (as the dominant ionization 
stage of Si atoms changes from Si~{\sc ii} to Si~{\sc iii}).

Therefore, the equivalent widths of Si~{\sc ii} 6347/6371 lines 
for each star were reanalyzed based on the corrected non-LTE calculations.
The resulting new values of $\Delta$ and $A^{\rm N}$ are shown 
in Figs.~11a and 11b, which should be compared with Figs.~4b and 4c of the
original article, As seen from these figures, while $\Delta$(old) values 
range between $\sim -0.4$ to $\sim 0.0$, $\Delta$(new)s are somewhat raised 
upward by $\sim 0.2$~dex on the average (i.e., ranging between $\sim -0.2$ 
and $\sim +0.2$). Since the gradual $T_{\rm eff}$-dependent effect 
is not so significant in the relevant range of 
$7000 \la T_{\rm eff} \la 13000$~K, the impact of applying new $\Delta$
is almost the overall raise of $A^{\rm N}$ (or [Si/H]) by $\sim 0.2$~dex,
which is not so important as compared to the star-to-star dispersion
of the abundances ($\sim 1$~dex). 
Accordingly, the main conclusion of the article (regarding the Si 
abundances of late A- to late B-stars) is not essentially affected by 
the revised non-LTE calculations.

In the meanwhile, the inadequate non-LTE calculations had a crucial 
influence upon the consequence of the Appendix of the paper, where the 
non-LTE effect on the formation of Si~{\sc ii} lines in B-type stars 
in general (covering $T_{\rm eff}$ up to $\sim 20000$~K) was 
passingly examined in comparison with Mashonkina's (2020, MNRAS, 
493, 6095) study, because the differences (increasing with $T_{\rm eff}$) 
become considerably large at such a high-$T_{\rm eff}$ regime.
This situation is illustrated in Fig.~12, which is the revised version of
the original Fig.~10. As shown in Fig.~12a, the degree of overionization 
($l_{0}^{\rm NLTE}/l_{0}^{\rm LTE} < 1$) is considerable and progressively 
enhanced with $T_{\rm eff}$ at $T_{\rm eff} \ga 10000$~K, while 
such a tendency was absent in the old Fig~10a. As a result, the behavior of
new $\Delta_{6371}$ (non-LTE correction for Si~{\sc ii} 6371; red solid line
in Fig.~12c) is markedly different as compared to the previous result (black 
dotted line in Fig.~12c); that is, it is larger by $\sim$~0.2--0.5~dex 
and turns into positive already around $T_{\rm eff} \sim$~13000~K. 

It was once stated in the Appendix that a reasonable non-LTE Si abundance 
could be obtained for the B3~IV star $\iota$~Her ($T_{\rm eff} \simeq 17500$~K)
due to an application of $\Delta_{6371} \sim -0.1$~dex, in contrast to  
Mashonkina's appreciably positive $\Delta_{6371}$ (+0.67) yielding an unacceptably 
large non-LTE Si abundance. However, this conclusion was wrong, because
such a slightly negative $\Delta_{6371}$ was fortuitously derived by incorrect
non-LTE calculations.  
The problem of an unreasonably high non-LTE Si abundance for $\iota$~Her 
from Si~{\sc ii} 6347/6371 lines still remains unsettled also in the author's 
non-LTE calculations. This means that much more investigation is further required 
towards correctly understanding the mechanism of Si~{\sc ii} line formation 
in B-type stars.

\setcounter{figure}{10}
\begin{figure}
\centerline{\includegraphics[width=0.5\textwidth,clip=]{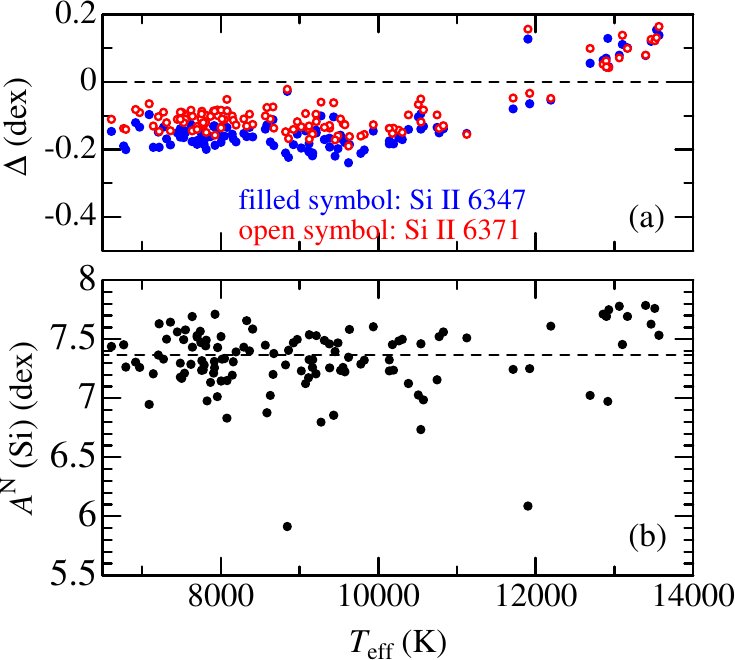}}
\caption{
(a) Non-LTE corrections for  Si~{\sc ii} 6347 ($\Delta_{6347}$, filled symbols) 
and Si~{\sc ii} 6371 ($\Delta_{6371}$, open symbols), plotted against $T_{\rm eff}$.
(b) $A^{\rm N}$(Si) (non-LTE Si abundance derived by averaging those of 
Si~{\sc ii} 6347/6371 lines), plotted against $T_{\rm eff}$.
Note that panels (a) and (b) are the revised Fig.~4b and Fig.~4c in the original
article.  
}
\label{fig1}
\end{figure}

\setcounter{figure}{11}
\begin{figure}
\centerline{\includegraphics[width=0.6\textwidth,clip=]{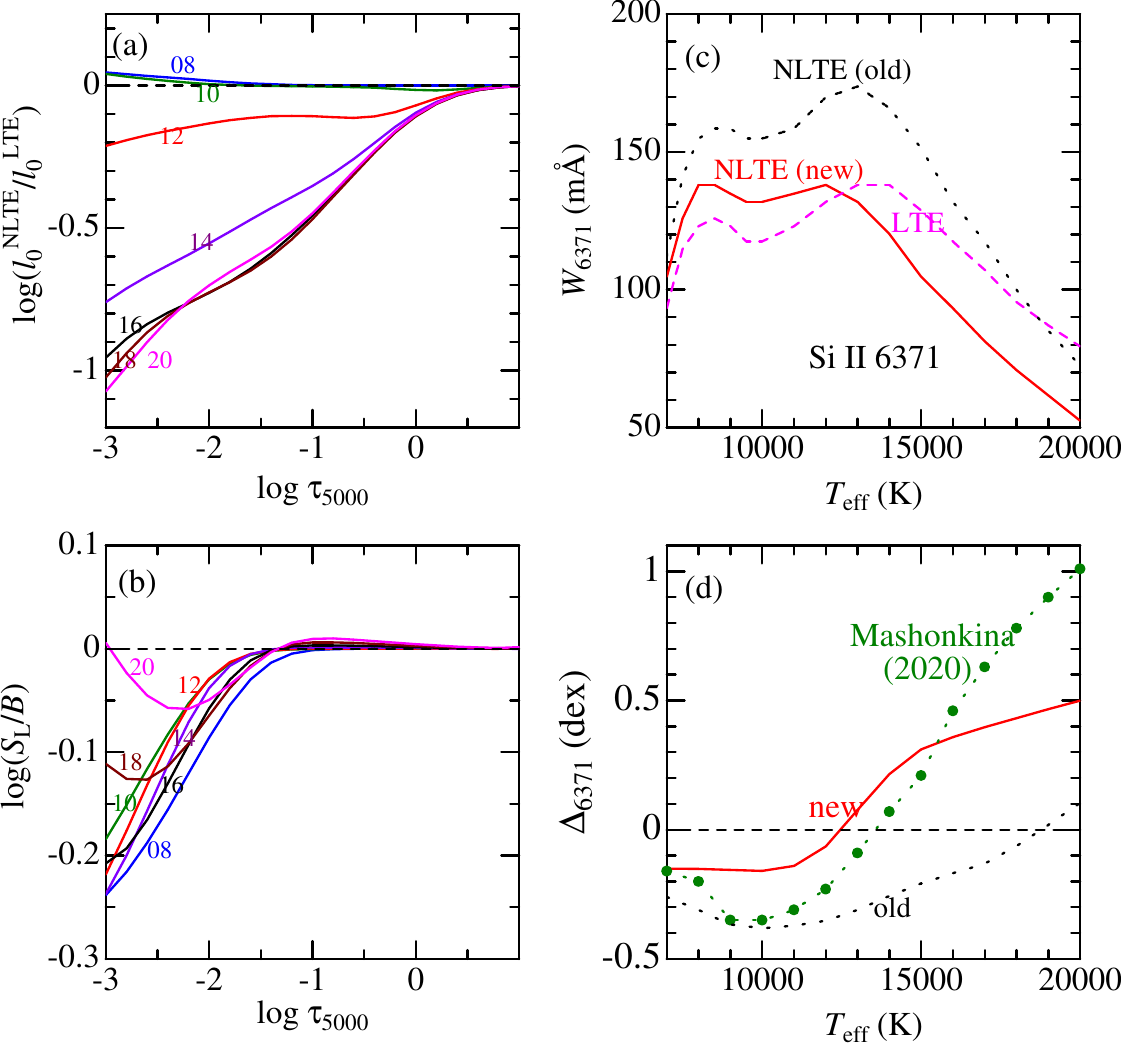}}
\caption{
Behaviors of the non-LTE effect on Si~{\sc ii} lines in B-type stars.  
This figure is the revised version of Fig.~10 in the original paper.
See the caption therein for more details. 
(a) The non-LTE-to-LTE line-center opacity ratio (vs. $\tau_{5000}$),  
(b) $S_{\rm L}/B$ ratio (vs. $\tau_{5000}$), 
(c) non-LTE and LTE equivalent widths for Si~{\sc ii} 6371 (vs. $T_{\rm eff}$),
and (d) non-LTE correction for Si~{\sc ii} 6371 (vs. $T_{\rm eff}$).
In panels (c) and (d), the old (wrong) results are also shown by black 
dotted lines for comparison.
}
\label{fig2}
\end{figure}


\begin{thebibliography}{}
\bibitem[]{}
  Aggarwal K. M., Keenan F. P., 2014, MNRAS, 442, 388
\bibitem[]{}
  Arenou F., Grenon M., G\'{o}mez A., 1992, A\&A, 258, 104
\bibitem[]{}
  Cunto W., Mendoza C., 1992, Rev. Mex. Astron. Astrofis., 23, 107
\bibitem[]{}
  ESA, 1997, The Hipparcos and Tycho Catalogues, ESA SP-1200, 
  available from NASA-ADC or CDS in a machine-readable form 
  (file name: hip\_main.dat)
\bibitem[]{} 
  Flower P. J., 1996, ApJ, 469, 355
\bibitem[]{}
  Ghazaryan S., Alecian G., 2016, MNRAS, 460, 1912
\bibitem[]{}
  Hoffleit D., Jaschek C., 1991,
  The Bright Star Catalogue, 5th revised ed. 
  (New Haven, Conn.: Yale University Observatory)
\bibitem[]{}
  Holweger H., St\"{u}renburg S., 1993, in Peculiar Versus Normal
  Phenomena in A-Type and Related Stars, ASP Conf. Ser. 44, 
  eds. M. M. Dworetsky, F. Castelli, and R. Faraggiana 
  (Astronomical Society of the Pacific: San Francisco), p.~356
\bibitem[]{}
  Kurucz R. L., 1993a, Kurucz CD-ROM, No. 13 (Harvard-Smithsonian Center
  for Astrophysics)
\bibitem[]{}
  Kurucz R. L., 1993b, Kurucz CD-ROM, No. 14 (Harvard-Smithsonian Center
  for Astrophysics)
\bibitem[]{}
  Kurucz R. L., Bell B., 1995, Kurucz CD-ROM, No. 23 
  (Harvard-Smithsonian Center for Astrophysics)
\bibitem[]{}
  Lejeune T., Schaerer D., 2001, A\&A, 366, 538
\bibitem[]{}
  Mashonkina L., 2020a, MNRAS, 493, 6095
\bibitem[]{}
  Mashonkina L., Ryabchikova T., Alexeeva S., Sitnova T., Zatsarinny, O.,  
  2020b, MNRAS, 499, 3706
\bibitem[]{}
  Michaud G., Alecian G., Richer J., 2015, Atomic Diffusion in Stars
  (Switzerland: Springer International Publishing)
\bibitem[]{}
  Napiwotzki R., Sch\"{o}nberner D., Wenske, V., 1993,
  A\&A, 268, 653
\bibitem[]{}
  Niemczura E., Murphy S. J., Smalley B., Uytterhoeven K., Pigulski A.,
  Lehmann H., Bowman D. M., Catanzaro G., van Aarle E., et al., 2015,
  MNRAS, 450, 2764
\bibitem[]{}
  Preston G. W., 1974, ARA\&A, 12, 257
\bibitem[]{}
  Richer J., Michaud G., Turcotte S., 2000, ApJ, 529, 338
\bibitem[]{}
  Saffe C., Miquelarena P., Alacoria J., Flores M., Jaque Arancibia M.,
  Calvo D., Mart\'{\i}n Girardi G., Grosso M., Collado A., 2021, A\&A, 647, A49
\bibitem[]{}
  Takeda Y., 1991, A\&A, 242, 455
\bibitem[]{}
  Takeda Y., 1995, PASJ, 47, 287
\bibitem[]{}
  Takeda Y., Han I., Kang D.-I., Lee B.-C., Kim K.-M., 2008,
  JKAS, 41, 83 
\bibitem[]{}
  Takeda Y., Kambe E., Sadakane K., Masuda S., 2010, PASJ, 62, 1239
\bibitem[]{}
  Takeda Y., Kang D.-I., Han I., Lee B.-C., Kim K.-M., 2009,
  PASJ, 61, 1165 
\bibitem[]{}
  Takeda Y., Kang D.-I., Han I.,  Lee B.-C., Kim K.-M.,
  Kawanomoto S., Ohishi N., 2012, PASJ, 64, 38
\bibitem[]{}
  Takeda Y., Kawanomoto S., Ohishi N., 2007, PASJ, 59, 245
\bibitem[]{}
  Takeda Y., Kawanomoto S., Ohishi N., 2014, PASJ, 66, 23
\bibitem[]{}
  Takeda Y., Kawanomoto S., Ohishi N., Kang D.-I., Lee B.-C., 
  Kim K.-M., Han I., 2018, PASJ, 70, 91 (T18)
\bibitem[]{}
  Takeda Y., Ohkubo M., Sato B., Kambe E., Sadakane K., 2005b, PASJ, 57, 27
\bibitem[]{}
  Takeda Y., Sadakane K., 1997, PASJ, 49, 367
\bibitem[]{}
  Takeda Y., Sato B., Kambe E., Masuda S., Izumiura H., Watanabe E., 
  Ohkubo M., et al., 2005a, PASJ, 57, 13
\bibitem[]{}
  Talon S., Richard O., Michaud G., 2006, ApJ, 645, 634
\bibitem[]{}
  van Leeuwen F., 2007, Hipparcos, the New Reduction of the Raw Data,
  Astrophysics and Space Science Library, Vol. 350
  (Berlin: Springer)
\end{thebibliography}
\end{document}